\documentclass{aa}
\usepackage{graphicx}
 \usepackage{color}
\usepackage{subfigure}

\usepackage{color}

\begin{document}      

   \title{The flaring H{\sc i} disk of the nearby spiral galaxy NGC~2683\thanks{Based on NRAO VLA observations (AI134).
The National Radio Astronomy Observatory is a facility of the National Science Foundation operated under cooperative 
agreement by Associated Universities, Inc.}}

   \author{B.~Vollmer, F.~Nehlig, R.~Ibata}

   \offprints{B.~Vollmer, e-mail: Bernd.Vollmer@astro.unistra.fr}

   \institute{Observatoire astronomique de Strasbourg, UMR 7750, 11, rue de l'universit\'e,
              67000 Strasbourg, France 
              }

   \date{Received / Accepted}

   \authorrunning{Vollmer, Nehlig, Ibata}
   \titlerunning{The flaring H{\sc i} disk of NGC~2683}

\abstract{
New deep VLA D array H{\sc i} observations of the highly inclined nearby spiral galaxy NGC~2683 are presented.
Archival C array data were processed and added to the new observations.
To investigate the 3D structure of the atomic gas disk, we made different 3D models for which 
we produced model H{\sc i} data cubes. The main ingredients of our best-fit model are 
(i) a thin disk inclined by $80^{\circ}$; (ii) a crude approximation of a spiral and/or bar structure by an 
elliptical surface density distribution of the gas disk; (iii) a slight warp in inclination between $10$~kpc$ \leq R \leq 20$~kpc (decreasing by 10$^{\circ}$);
(iv) an exponential flare that rises from $0.5$~kpc at $R=9$~kpc 
to $4$~kpc at $R=15$~kpc, stays constant until $R=22$~kpc, and decreases its height for $R > 22$~kpc; and (v) 
a low surface-density gas ring with a vertical offset of $1.3$~kpc.
The slope of NGC~2683's flare is comparable, but somewhat steeper than those of other spiral galaxies. 
NGC~2683's maximum height of the flare is also comparable to those of other galaxies. On the other hand,
a saturation of the flare is only observed in NGC~2683.
Based on the comparison between the high resolution model and observations, we exclude the existence of 
an extended atomic gas halo around the optical and thin gas disk.
Under the assumption of vertical hydrostatic equilibrium we derive the vertical velocity dispersion of the gas.
The high turbulent velocity dispersion in the flare can be explained by energy injection by (i) supernovae,
(ii) magneto-rotational instabilities, (iii) ISM stirring by dark matter substructure, or (iv)
external gas accretion. The existence of the complex large-scale warping and asymmetries favors external gas accretion as
one of the major energy sources that drives turbulence in the outer gas disk.
We propose a scenario where this external accretion leads to turbulent adiabatic compression that enhances
the turbulent velocity dispersion and might quench star formation in the outer gas disk of NGC~2683.
}

\maketitle

\section{Introduction \label{sec:intro}}

The neutral gas content of a spiral galaxy can be roughly divided into three components: the dense
molecular disk, the atomic gas disk, and an atomic gas halo. Whereas the inner gas disk is extremely thin
($\sim 100$~pc; e.g. Kalberla \& Kerp 2009 for the Galaxy), 
the gas halo component, if present, is much more extended 
(a few kpc up to $22$~kpc; Gentile et al 2013, Oosterloo et al. 2007)
and contains about 15\,\% of the galaxy's total gas content (NGC~891, Oosterloo et al. 2007; NGC~6946, 
Boomsma et al. 2008; NGC~253, Boomsma et al. 2005; NGC~4559, Barbieri et al. 2005; UGC~7321, 
Matthews \& Wood 2003; NGC~2403, Fraternali et al. 2002; NGC~2613, Chaves \& Irwin 2001; NGC~3198, Gentile et al. 2013; 
for a review see Sancisi et al. 2008). 
In edge-on systems, like NGC~891, the vertical extension of the halo gas is directly accessible. 

The halo gas typically shows differential rotation parallel to the disk plane. Compared to the disk, 
the halo gas rotation velocity is lower (Fraternali et al. 2002, Oosterloo et al. 2007).
In some cases, radial gas inflow is detected (Fraternali et al. 2002, Zschaechner et al. 2012). 
Above the star forming disk, the ionized component of the gas halo is detected in H$\alpha$
(e.g., Rossa \& Dettmar 2003), radio continuum (e.g., Dahlem et al. 2006), and X-rays 
(e.g., Yamasaki et al. 2009). 

It is not clear what role the halo gas plays in the ecosystem of a spiral galaxy:
\begin{itemize}
\item
Above the star forming disk, galactic fountains (Shapiro \& Field 1976, Fraternali \& Binney 2006) may play a 
dominant role. The disk gas is lifted into the halo by stellar winds and supernova explosions; then it cools there and
finally falls back onto the galactic disk.   
\item
The gas lifted by galactic fountains may interact with a pre-existing hot ionized halo, leading
to enhanced gas condensation and gas backfall.
\item
Accretion of intergalactic gas may add gas to the halo component. This accretion can be in form of the merging
of small gas-rich satellites (van der Hulst \& Sancisi 2005).  
\end{itemize}
The last scenario is corroborated by the detection of faint gaseous tidal streams in a considerable 
number of local spiral galaxies (Sancisi et al. 2008). 

The H{\sc i} disk of spiral galaxies is often warped and flares beyond the optical radius; i.e., its thickness increases
exponentially. Flares are detected in the Galaxy (Kalberla \& Kerp 2009), M~31 (Brinks \& Burton 1984),
and a sample of edge-on spiral galaxies (O'Brien et al. 2010, Zschaechner et al. 2012). While warps are ubiquitous in spiral galaxies
where the H{\sc i} disk is more extended than the optical disk (Garcia-Ruiz et al. 2002),
flaring gas disks are common, but not ubiquitous: 
for modeling the H{\sc i} observations of NGC~5746 (Rand \& Benjamin 2008) and NGC~4559 (Barbieri et al. 2005), 
a flaring gas disk is not necessary.
According to van der Kruit (2007), the inner flat disk and the outer warped (and in some cases flared) one are distinct components 
with different formation histories. The inner disk forms initially, and the warped outer disk forms as a result of much later infall of
gas with a higher angular momentum in a different orientation.

NGC~2683 is a good candidate for further investigating the role of the gas disk beyond the optical disk and gas halos for the evolution of a spiral galaxy. Its basic parameters are presented in Table~\ref{tab:param}. 
This highly inclined spiral galaxy is located at a distance of 7.7~Mpc. Within the galaxy a distance of $1$~kpc thus 
corresponds to $27''$. Deep VLA D array observations with a resolution of $\sim 1'$ are ideal for probing its
neutral gas halo. Casertano \& van Gorkom (1991) observed NGC~2683 for one hour with the VLA in D array
configuration. They found neutral hydrogen extending more than twice as far as the visible light on both sides of
the galaxy. The gas distribution is fairly symmetric and close to the plane of the optical disk.
The derived rotation curve peaks at about 215~km\,s$^{-1}$ at $\sim 3$~kpc from the galaxy center and
then decreases monotonically. Kuzio de Naray et al. (2009) found a multi-valued, figure-of-eight velocity
structure in the inner $45''=1.7$~kpc of the long-slit spectrum of NGC~2683 and twisted isovelocity contours in the velocity field. 
They argue that these features, along with boxy galaxy isophotes, are evidence of a bar in NGC 2683.

In this article we revisit NGC~2683, which we observed for nine hours in the H{\sc i} 21cm line
with the VLA in D configuration. We also reduced archival C array observations and added them to
our data.
The observations are presented in Sec.~\ref{sec:observations}, and
the results are given in Sec.~\ref{sec:results}. The 3D kinematically modeling of the galaxy is described in 
Sec.~\ref{sec:3D} followed by the modeling results. We discuss our findings in Sec.~\ref{sec:discussion} and
give our conclusions in Sec.~\ref{sec:conclusions}.

\begin{table}
      \caption{Basic parameters of NGC~2683}
         \label{tab:param}
         \begin{tabular}{lll}
           \hline
           Type & Sb \\
           m$_{\rm B}^{0}$ & $9.84$~mag & (de Vaucouleurs et al. 1976)\\
           $D_{25}$ & $9.3'=20.8$~kpc & (Nilson 1973) \\
           Distance & $7.7$~Mpc & (Tonry et al. 2001)\\
           M$_{\rm B}$ & $-19.59$~mag & \\
           $v_{\rm rot}^{\rm max}$ & $215$~km\,s$^{-1}$ & (Casertano \& van Gorkom 1991) \\ 
           $\dot{M}_{*}$ & 0.8~M$_{\odot}$yr$^{-1}$ & (Irwin et al. 1999) \\
           $l_{\rm B}$ & $0.81'=1.8$~kpc & (Kent 1985) \\
           $l_{\rm K'}$ & $0.67'=1.5$~kpc & (this paper) \\
           $M_{*}$ & $3.6 \times 10^{10}$~M$_{\odot}$  & (this paper) \\
           $M_{\rm *,disk}$ & $2.6 \times 10^{10}$~M$_{\odot}$ & (this paper) \\
           \hline
         \end{tabular}
\end{table}

\section{Observations \label{sec:observations}}

NGC~2683 was observed in December 2009 for nine hours with the Very Large Array (Napier et al. 1983) in
D array configuration. 
We used 0542+498 as flux calibrator. The phase calibrator 0909+428 was observed for 2.5min every 30min.
The total bandwidth of 3.125~MHz was divided into 128 channels with a
channel separation of 5.16~km\,s$^{-1}$. Of the 22 available antennas, 20 were EVLA antennas. 
To avoid closure errors on the VLA-EVLA baselines, an alternative 'channel 0' was created  for the
initial calibration stage.
Subsequent calibration was achieved using VLA standard calibration procedures.
Fifteen line-free channels were selected for continuum subtraction.
The maps were generated with natural weighting to maximize sensitivity.
We obtained an rms noise level of 1~mJy/beam in a 5~km\,s$^{-1}$ channel.
The resolution is $61'' \times 51''$.

In addition, we reduced 8.5~h of archival C array data.
The total bandwidth of 3.125~MHz was divided into 64 channels with a
channel separation of 10.3~km\,s$^{-1}$. Calibration was done using VLA standard calibration procedures.
The maps were generated with a weighting between natural and uniform (ROBUST=1).
An rms noise level of 0.4~mJy/beam in a 10.3~km\,s$^{-1}$ channel was obtained with a
resolution of $19'' \times 18''$.

In a last step, the C and D array observations were combined into a single data set.
With a ROBUST=1 weighting leading to a resolution of $21'' \times 20''$, we obtained
an rms noise level of 0.3~mJy/beam assuming a linewidth of 10.3~km\,s$^{-1}$ channel. 
The cube was clipped at a flux density of 2.0~mJy/beam or $7$ times the rms.
Linear fits were made to the line-free channels of each spectrum.
These baselines were then subtracted from the spectra.
The H{\sc i} column densities corresponding to the rms noise levels in individual channel maps are given in
Table~\ref{tab:paramhiobs}.
\begin{table*}
      \caption{H{\sc i} observations and the combined data cube}
         \label{tab:paramhiobs}
         \begin{tabular}{llll}
           \hline
           Date & December 2009 & May 2004 & \\
           Array & D & C & CD \\
           Duration & 9h & 8.5h & 17.5h \\
           Bandwidth & 3.125~MHz & 3.125~MHz & 3.125~MHz \\
           Number of channels & 128 & 64 & 64 \\
           Central velocity$^{\rm a}$ & 412~km\,s$^{-1}$ & 412~km\,s$^{-1}$ & 412~km\,s$^{-1}$ \\
           Velocity resolution & 5.16~km\,s$^{-1}$ & 10.30~km\,s$^{-1}$ & 10.30~km\,s$^{-1}$ \\
           FWHP restoring beam & $61'' \times 51''$ & $19'' \times 18''$ & $21'' \times 20''$ \\
           rms noise in channel maps &  1.0~mJy/beam & 0.4~mJy/beam & 0.3~mJy/beam\\
           Equivalent H{\sc i} column density & $2 \times 10^{18}$~cm$^{-2}$ & $1.3 \times 10^{19}$~cm$^{-2}$ & $8 \times 10^{18}$~cm$^{-2}$ \\
           \hline
         \end{tabular}
       
         $^{a}$ heliocentric, optical definition 
\end{table*}
For the creation of the moment0 map, a boxcar smoothing of $7 \times 7 \times 7$ pixels ($49'' \times 49'' \times 36$~km\,s$^{-1}$) was applied
to the data cube and voxels of the smoothed cube with flux densities lower than $1/0.3$~mJy/beam for the D/C+D array data were blanked.
The same voxels were then blanked in the original data cube before generating the moment0 map.
The primary beam correction was then applied to the moment0 maps. At the end, the primary beam correction was applied to the channel maps.
For the comparison with our models, we smoothed the D array data 
cube to the spectral resolution of the C+D array data cube ($10.3$~km\,s$^{-1}$).

We made continuum maps from the line-free channels of the H{\sc i} data cube.
However, owing to the small number of line-free channels, our 20cm continuum image has a high
rms noise level of $0.35$~mJy/beam at a resolution of $60''$ compared to the rms noise level of $1$~mJy/beam 
in a $5.16$~km\,s$^{-1}$ channel (Table~\ref{tab:paramhiobs}). 
On this image only the inner star forming disk of $\sim 4'$ diameter is visible.

\section{Results \label{sec:results}}

\begin{figure}
        \resizebox{\hsize}{!}{\includegraphics{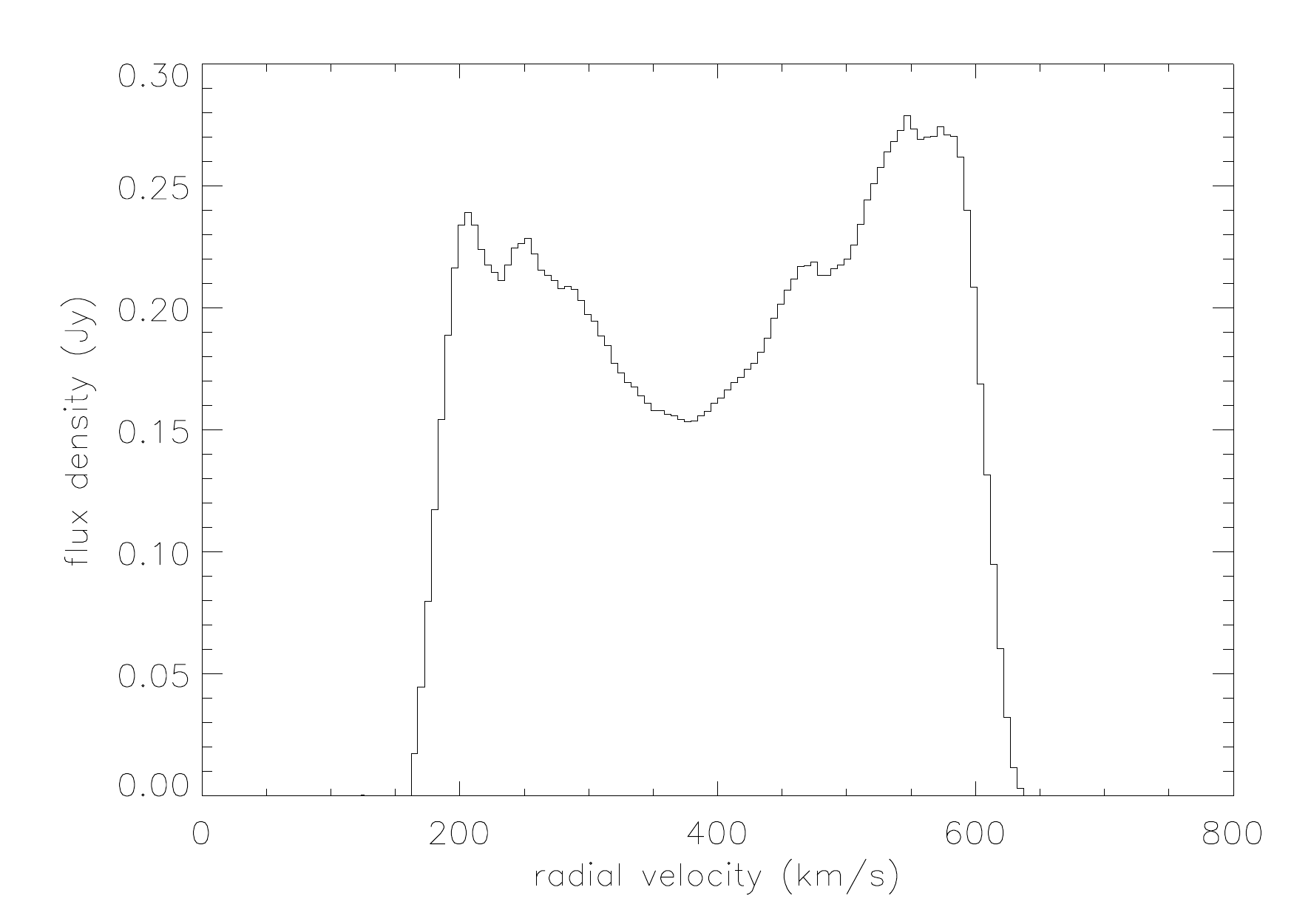}}
        \caption{Integrated H{\sc i} spectrum of NGC~2683 from the VLA D array observations.
        } \label{fig:spectrum}
\end{figure}

We extracted an integrated spectrum from the area containing line emission (Fig.~\ref{fig:spectrum}).
The total line flux is $F_{\rm HI}=101.4$~Jy\,km\,s$^{-1}$.  
This is coincidentally equal to the line flux that is corrected for self-absorption, but higher
than the pointing/extension-corrected line flux of $F_{\rm HI}=87.6$~Jy\,km\,s$^{-1}$ from Springob et al. (2005).
With a distance of 7.7~Mpc, this corresponds
to a total H{\sc i} mass of $M_{\rm HI}=1.42 \times 10^{9}$~M$_{\odot}$.
The line flux of the C array data is $F_{\rm HI}=73.8$~Jy\,km\,s$^{-1}$ corresponding to 
$M_{\rm HI}=1.03 \times 10^{9}$~M$_{\odot}$.
The integrated line profile is asymmetric with a 16\,\% higher peak flux density on the receding side.
The linewidths at 20\,\% and 50\,\% of the peak flux density are $W_{20}=450$~km\,s$^{-1}$ and $W_{50}=426$~km\,s$^{-1}$.
With a systemic velocity of $415$~km\,s$^{-1}$ (Casertano \& van Gorkom 1991), the approaching and receding sides of the galactic
disk contain equal amounts of atomic gas. 
The channel maps of the D array and combined C+D array observations are presented in Appendix~\ref{app}.

\subsection{H{\sc i} gas distribution \label{sec:hidistribution}}

\begin{figure*}
        \resizebox{\hsize}{!}{\includegraphics{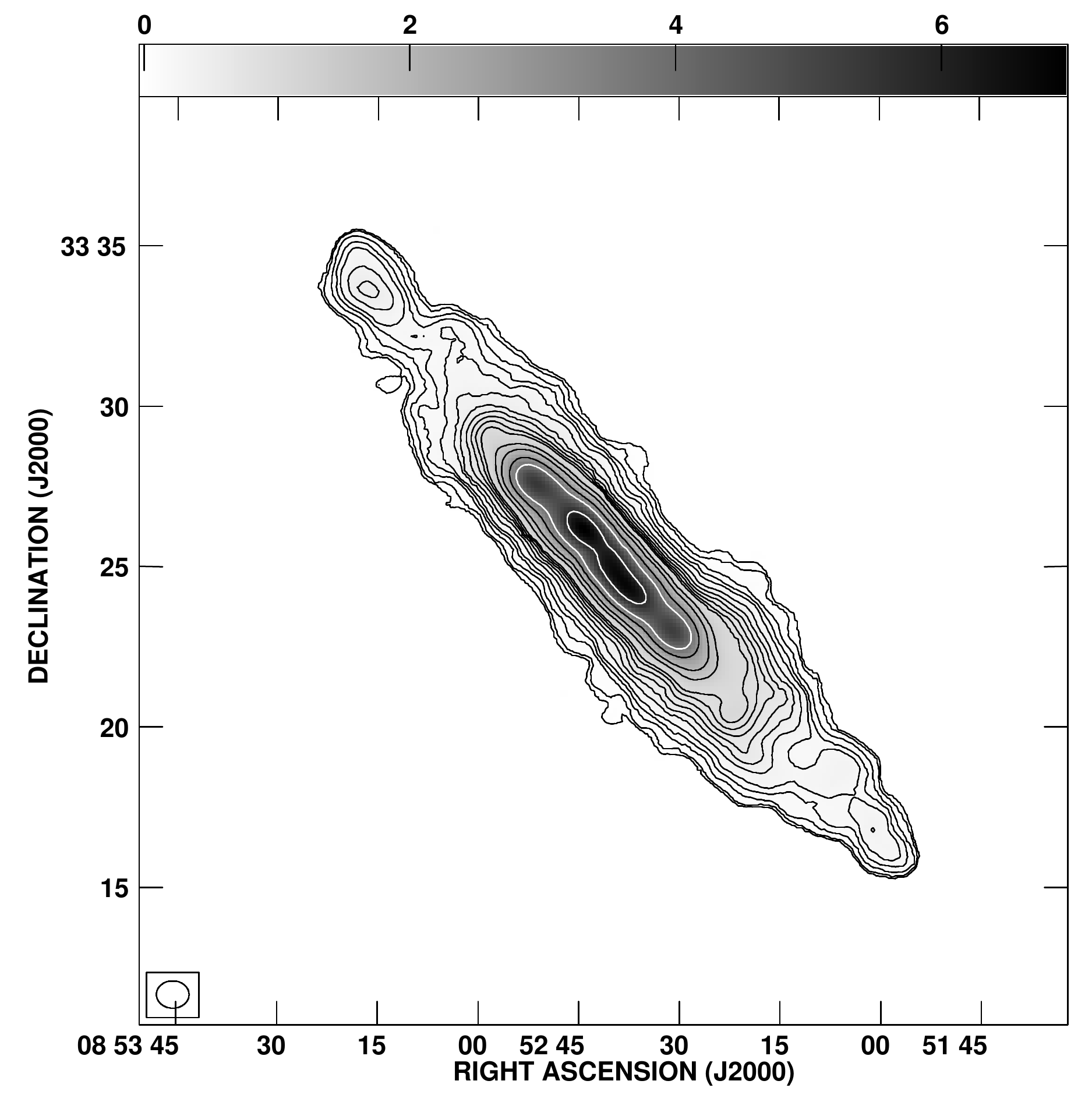}\includegraphics{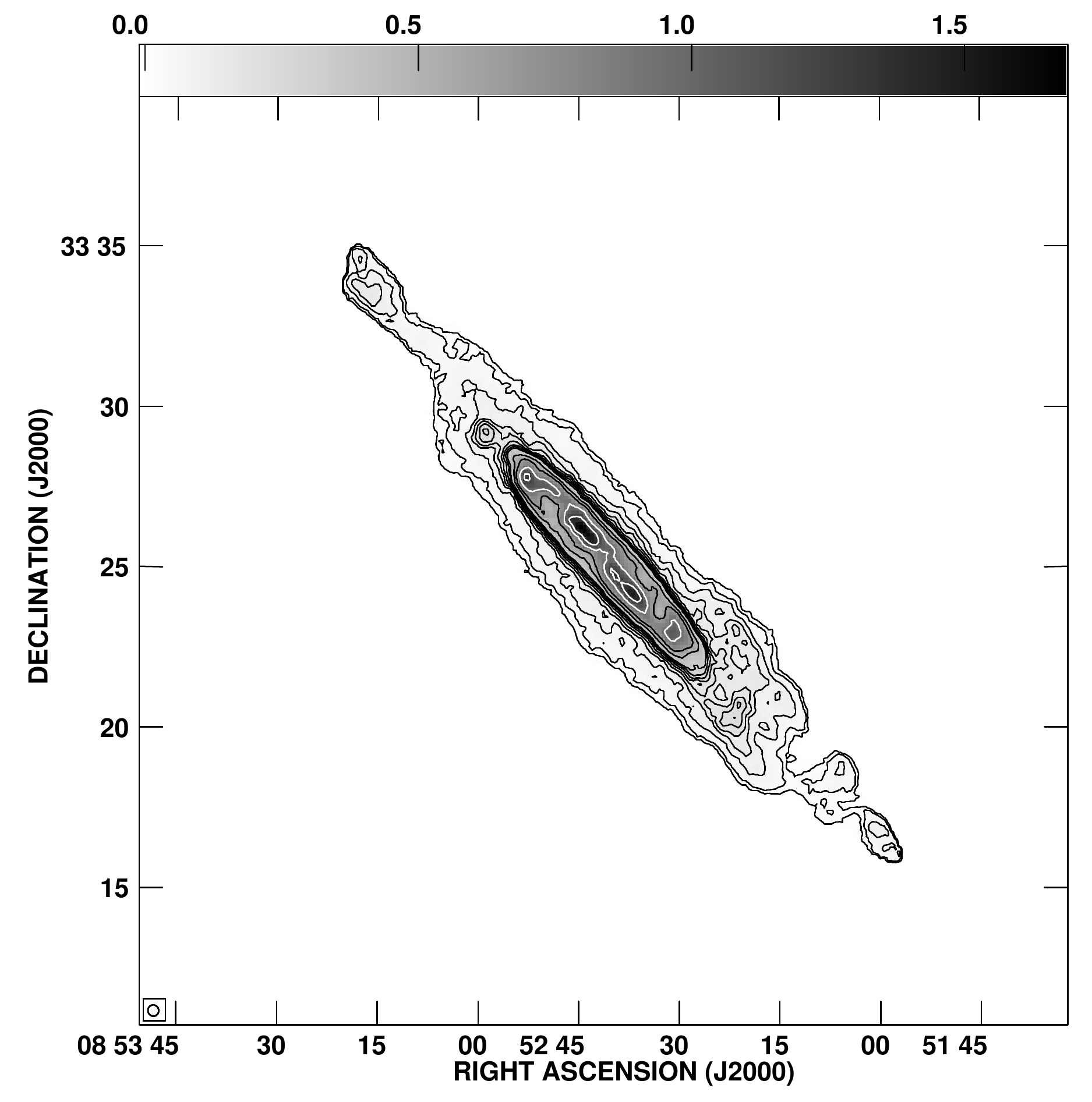}}
        \caption{H{\sc i} gas distribution of NGC~2683. Left panel: from the D array observations. The beam size is
          $61'' \times 51''$. The contour levels are (1,2,4,6,8,12,16,20,24,28,32,48,64,96,128,192,264,392)
          $\times$ $30$~mJy/beam\,km\,s$^{-1}$ or $1.1 \times 10^{19}$~cm$^{-2}$. 
          The horizontal wedge is in units of Jy/beam\,km\,s$^{-1}$. Right panel:
          from the C+D array observations. The beam size is
          $21'' \times 20''$. The contour levels are (2,4,6,8,12,16,20,24,28,32,48,64,96,128,192,264,392)
          $\times$ $10$~mJy/beam\,km\,s$^{-1}$ or $2.6 \times 10^{19}$~cm$^{-2}$.
          Grayscales are in units of Jy/beam\,km\,s$^{-1}$. As in Casertano \& van Gorkom (1991)
          we deliberately have suppressed the enhanced noise due to the primary beam correction (for comparison see
          Figs.~\ref{fig:PV2BR}, \ref{fig:PV2HR}, \ref{fig:channels_D}, and \ref{fig:channels_CD}).
        } \label{fig:ngc2683_d_mom0_1}
\end{figure*}

\begin{figure}
        \resizebox{\hsize}{!}{\includegraphics{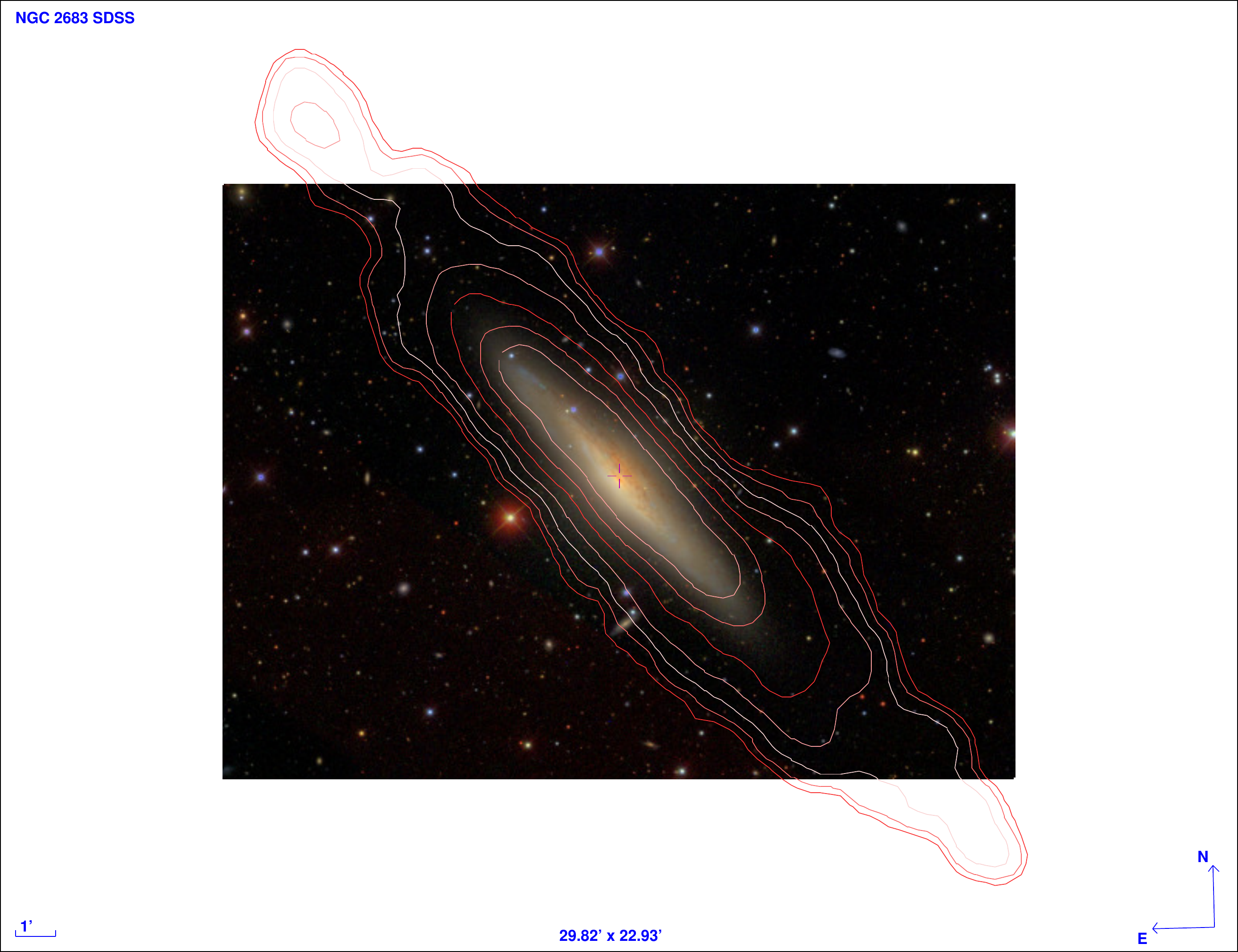}}
        \caption{H{\sc i} contours on the SDSS color image of NGC~2683 (copyright 2006 Michael R. Blanton \& 
          David W. Hogg). The contour levels are (2,4,8,16,32,64,128)$\times$ $10$~mJy/beam\,km\,s$^{-1}$ or $2.6 \times 10^{19}$~cm$^{-2}$.
        } \label{fig:SDSS}
\end{figure}

The distribution of atomic hydrogen of NGC~2683 is presented in Fig.~\ref{fig:ngc2683_d_mom0_1}.
As already shown by Casertano \& van Gorkom (1991), the atomic hydrogen is distributed over a diameter of 
$26.5'$,so almost three times the optical diameter (Fig.~\ref{fig:SDSS}).
The overall structure is that of an inner disk of high column density gas ($\sim 9'$ diameter) and
an extended low column-density vertically extended structure. We show that a part of this vertical extent
is caused by the projection of a flaring gas disk.
While the high column gas disk is thin, the low column density gas is more extended in the vertical direction.

The thin gas disk appears significantly bent. Whereas the southwestern part of the high column density gas disk
at a distance of $3.5'$ from the galaxy center bends away from the major axis to the south, the northeastern 
part bends slightly to the north. This warping is reversed at distances between $3.5'$ and $8'$
from the galaxy center, which means that at these distances the northeastern and southwestern parts bend
toward the major axis. The projected disk thickness increases slightly with increasing distance from the
galaxy center. At distances greater
than $8'$, the vertical projected thickness of the gas disk decreases rapidly. At the extremities of the gas disk, at
distances of $\sim 11.5'=26$~kpc, two distinct gas blobs are present. The northeastern blob has a
column density that is twice as high as the southwestern one. In the higher resolution C+D array image, the southwestern part of the 
gas disk (08 52 20, +33 22 00, red circle in the right panel of Fig.~\ref{fig:ngc2683_d_mom0_1}) 
has a significant substructure. The gas mass within the depicted area is approximately $6 \times 10^{7}$~M$_{\odot}$. 
The mass excess compared to the opposite side of the galactic disk (08 52 55, +33 30 00,  a second red circle in the 
right panel of Fig.~\ref{fig:ngc2683_d_mom0_1}) is about $1.5 \times 10^{7}$~M$_{\odot}$.

\subsection{H{\sc i} kinematics \label{sec:hikinematics}} 

\begin{figure}
        \resizebox{\hsize}{!}{\includegraphics{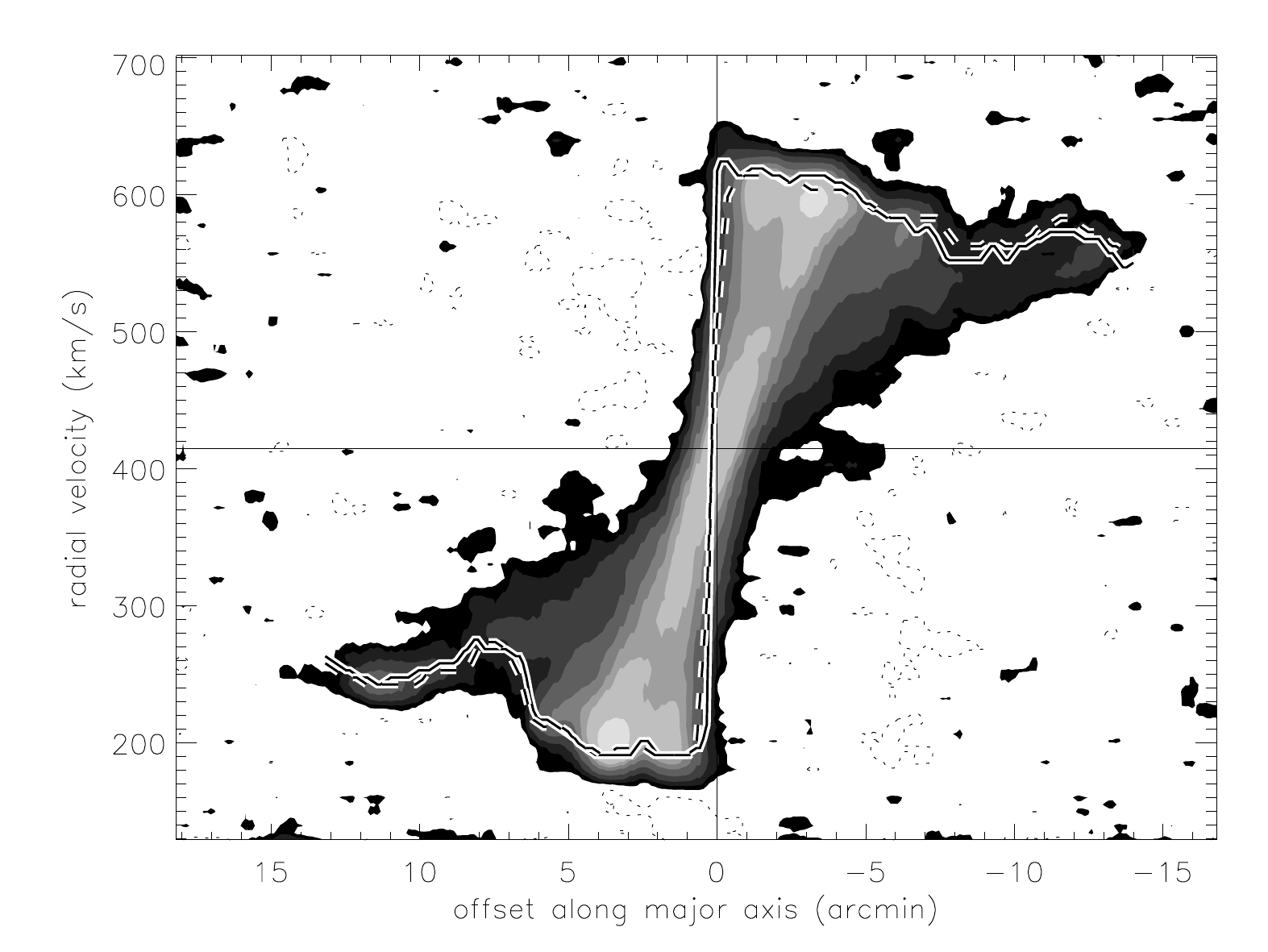}}
        \caption{Position velocity diagram along the major axis from the D array observations.
          At each position along the major axis, the highest velocities trace the rotation curve at that radius
            (solid line). The dashed line shows the rotation velocity derived with the envelope-tracing method. 
          The grayscale levels are $(-1,1,2,4,8,16,24,32,64,132) \times 1$~mJy/beam. 
        } \label{fig:rotcurve}
\end{figure}

The velocity field of highly inclined galaxies cannot be used to derive the kinematical parameters 
(e.g., Kregel \& van der Kruit 2004) that have to be derived directly from the data cube.
We derived the rotation curve of NGC~2683 from the position velocity diagram along the major axis and validated 
it with our 3D model (Sect.~\ref{sec:3D}).
To search for counter-rotating gas and determine the rotation curve, 
we decreased the rms noise level in the position velocity diagrams by smoothing the velocity axis with a boxcar function of 
a width of three pixels or $15.5$~km\,s$^{-1}$.
The position velocity diagram along the major axis from the D array data is presented in Fig.~\ref{fig:rotcurve}. 
We did not observe any counter-rotating gas. 
To derive the rotation curve from the position velocity diagram, we assumed that the neutral hydrogen moves in
circular orbits and that the highest velocities trace the rotation curve at a given radius.
The envelope-tracing method (Sofue \& Rubin 2001) makes  use  of  the  terminal  velocity  in a position velocity  
diagram along the major axis. The rotation velocity $v_{\rm rot}$ is derived by using the terminal velocity $v_{\rm t}$:
\begin{equation}
v_{\rm rot}=(v_{\rm t}-v_{\rm sys})/\sin i - \Delta  \ ,
\end{equation}
where $\Delta=(\sigma_{\rm obs}^2+\sigma_{\rm ISM}^2)^{\frac{1}{2}}=11.3$~km\,s$^{-1}$ with $\sigma_{\rm ISM}=10$~km\,s$^{-1}$ and 
$\sigma_{\rm obs}=5.16$~km\,s$^{-1}$, which are the velocity dispersion of the interstellar medium and the velocity resolution of the 
observations, respectively. 

We adopted a similar approach:
the spectra along the major axis were decomposed into Gaussian profiles after clipping at a level of $2$~mJy/beam.
The maximum/minimum peak velocities of the Gaussian decompositions 
were then adopted as rotation velocities on the approaching/receding side of the galaxy and corrected for
an inclination of $83^{\circ}$ (see Sect.~\ref{sec:3D})\footnote{Up to $30$ Gaussian components were fitted to reproduce the spectra.
In the inner disk $15$-$25$ components were needed whereas in the outer disk $3$-$10$ components were sufficient to fit the H{\sc i} spectra.}.
Moreover, we set $\Delta=15.5$~km\,s$^{-1}$.
The derived velocity curve (solid line in Fig.~\ref{fig:rotcurve}) is very similar to those obtained with the envelope-tracing method (dashed line in Fig.~\ref{fig:rotcurve}) and derived by Casertano \& van Gorkom (1991; Fig.~2).

Position velocity diagrams parallel to the minor axis (positions are shown in the upper panels of Fig.~\ref{fig:mom0HR}) 
are presented in Figs.~\ref{fig:PV2BR} and \ref{fig:PV2HR}.
Almost all position velocity diagrams are asymmetric. In the inner gas disk, emission of higher surface brightness 
is found toward positive/negative offsets at the receding/approaching side of the galactic disk (p6/p4).
Low surface brightness emission of highest velocities is found toward positive offsets at the approaching side (pv4),
whereas the envelope of the emission distribution at the receding side is quite symmetric (pv6).
At intermediate radii of the receding side (pv7 and pv8) emission of highest surface brightness is found toward negative offsets. 
At intermediate radii of the approaching side (pv2 and pv3) most emission is found toward positive offsets.
In the outer part of the disk (pv1 and pv9), the emission is shifted toward positive offsets.

\section{3D modeling \label{sec:3D}}

\begin{figure}
  \subfigure{\resizebox{\hsize}{!}{\includegraphics{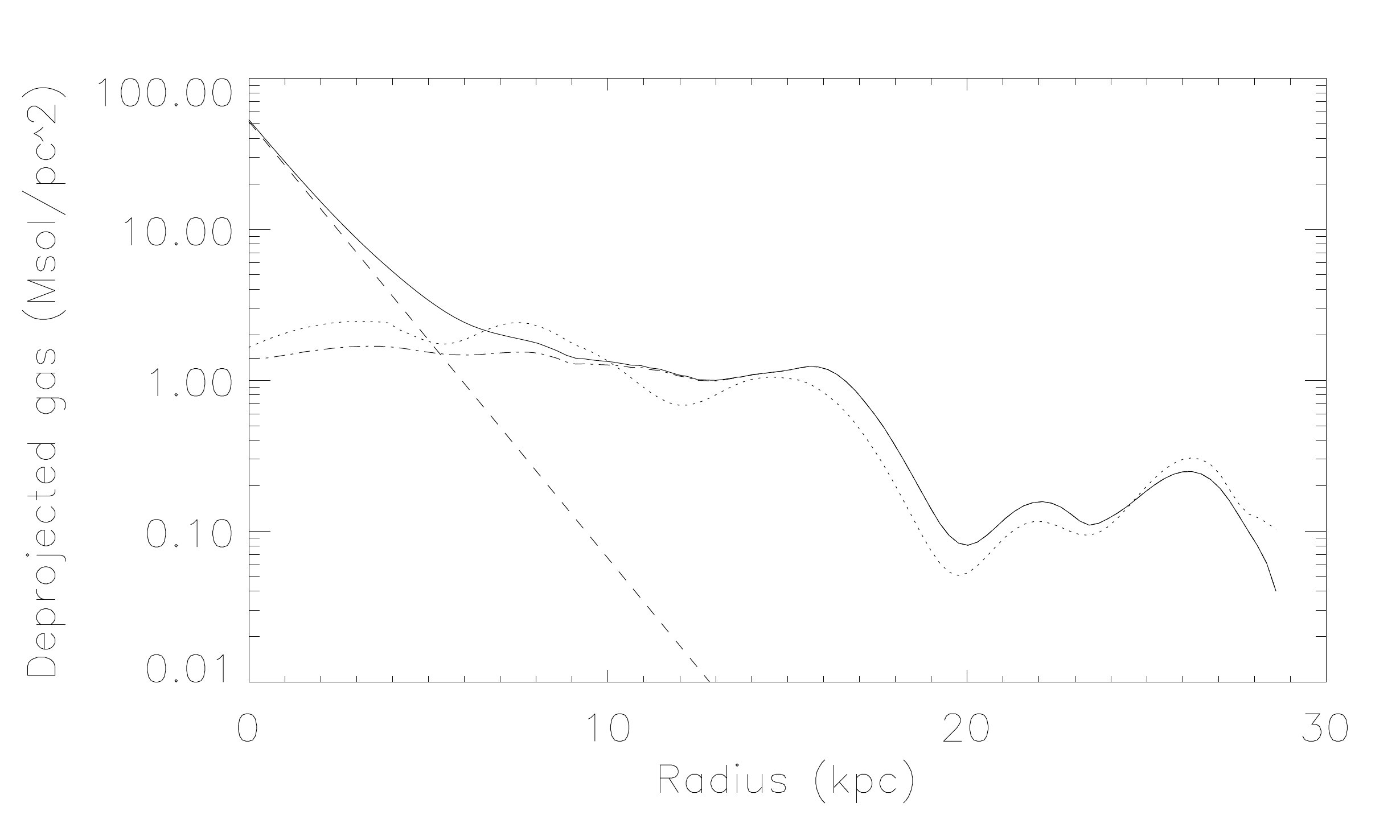}}}
  \subfigure{\resizebox{\hsize}{!}{\includegraphics{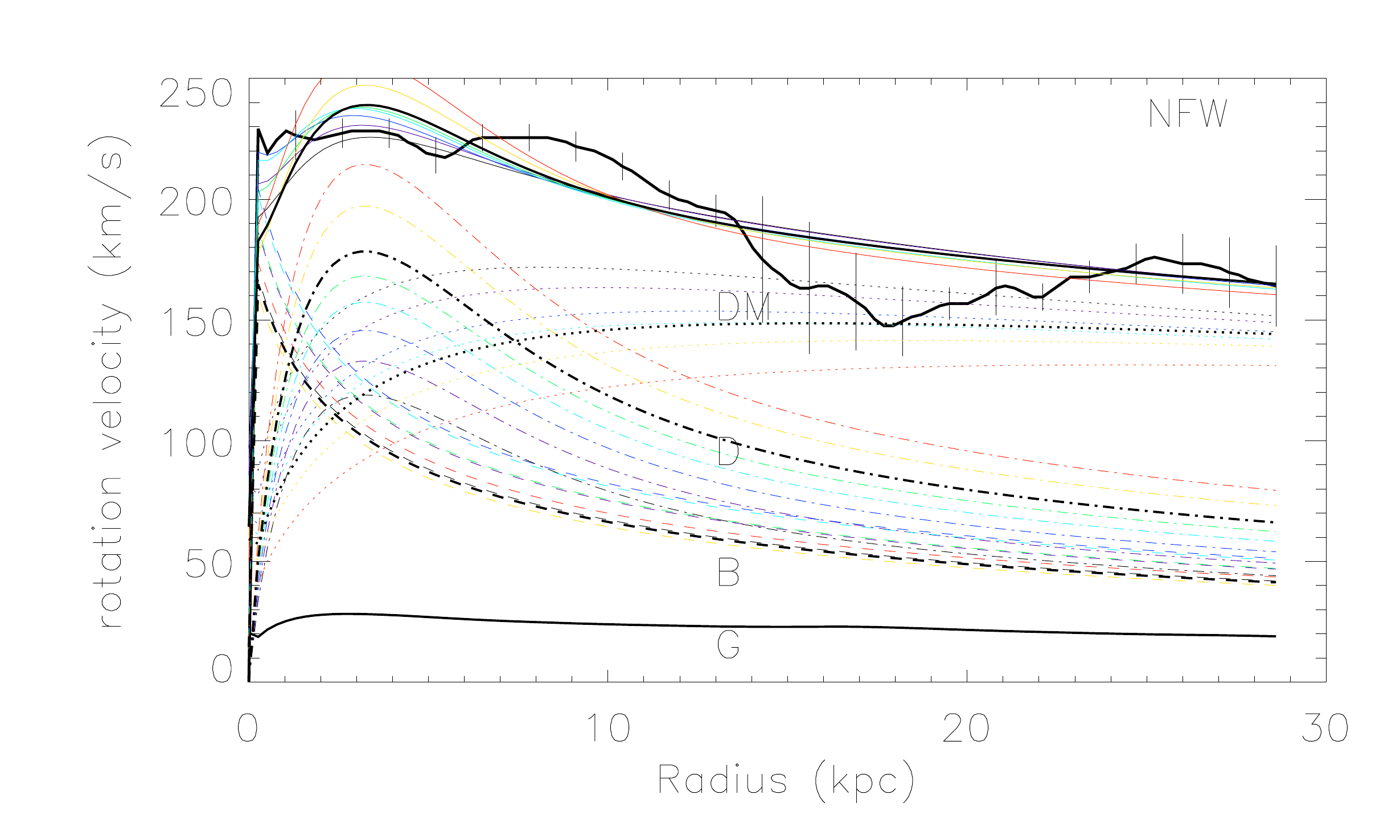}}}
  \subfigure{\resizebox{\hsize}{!}{\includegraphics{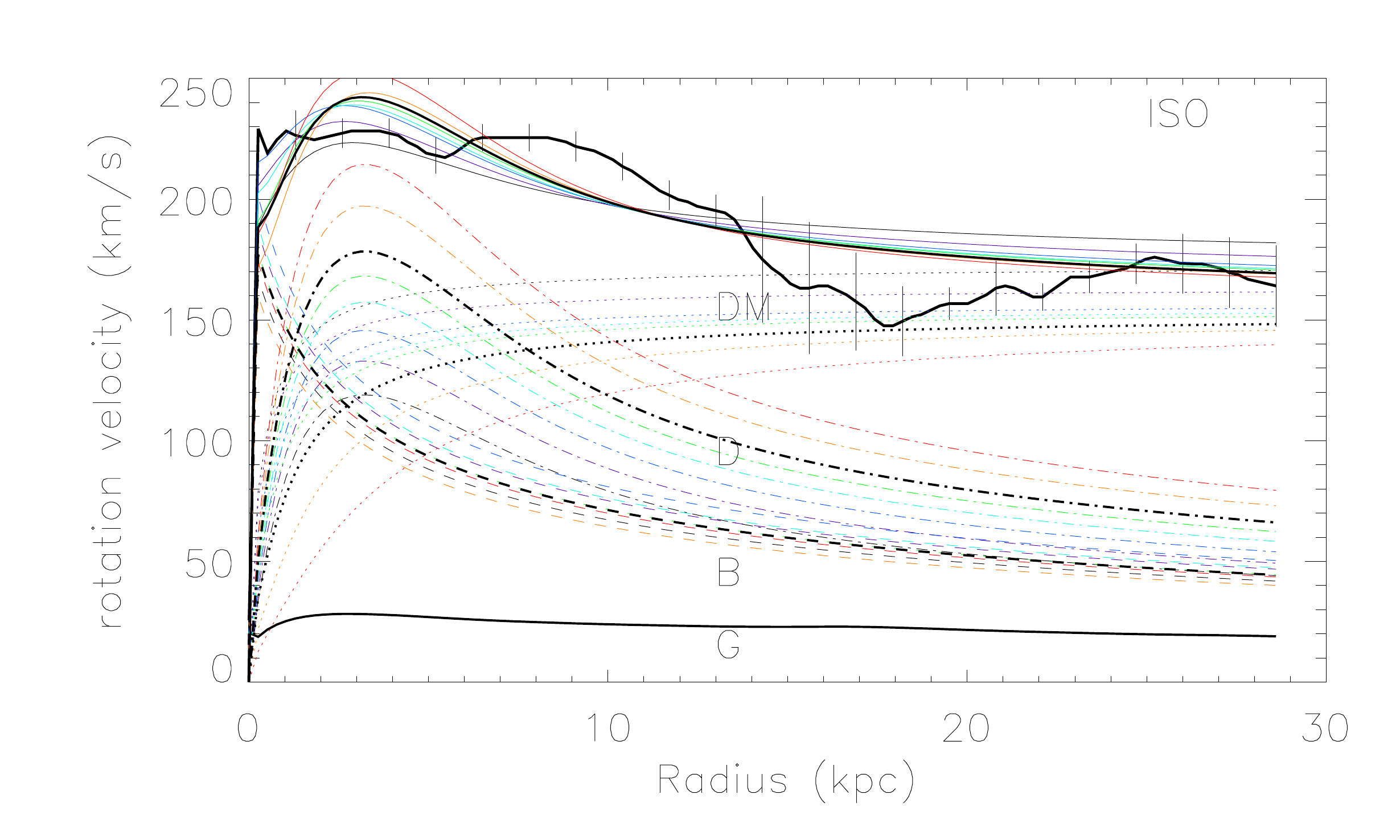}}}
        \caption{Radial profiles of NGC~2683. 
          Upper panel: deprojected total (solid), atomic (dash-dotted), and assumed
          molecular (dashed) gas surface density based on SEST CO-line observations (see Sect.~\ref{sec:3D}).
          The dotted line represents the symmetric deprojected H{\sc i} profile obtained from the method of Warmels (1988).
          Lower panel: observed and modeled rotation curve (solid),
          component due to the stellar bulge and disk (dashed), gas (dash-dotted), and dark matter halo (dotted).
          The colors indicate different K' band mass-to-light ratios from blue (0.5) to red (1.3).
        }\label{fig:deprojected_gas2683}
\end{figure} 

We developed an IDL code to construct 3D data cubes with properties
identical to the observed data cubes (D and C+D array data) from these model gas distributions.
The sensitivity limits of our observations were also applied to the model cubes.

For the 3D modeling we used a symmetrized mean rotation velocity (lower panel of Fig.~\ref{fig:deprojected_gas2683}),
which we obtained by the Gaussian decomposition of the spectra along the major axis corrected by inclination
(lower panel of Fig.~\ref{fig:deprojected_gas2683}).
The rotation curve is consistent with the one derived by Casertano \& van Gorkom (1991). It is flat ($220$~km\,s$^{-1}$) up
to a galactic radius of $\sim 10$~kpc. The rotation velocity declines to a value of $160$~km\,s$^{-1}$ at $18$~kpc
and then rises to $200$~km\,s$^{-1}$ at a galactic radius of $26$~kpc.

Following Warmels (1988), we derived the deprojected H{\sc i} surface density using the symmetric
rotation curve and the major axis position velocity diagram.
First, the so-called strip integral was calculated by integrating the H{\sc i} column
density map perpendicularly to the major axis, as defined by the
dynamical center and position angle of the stellar disk. The approaching and the receding sides of this strip integral
are deprojected separately, using the iterative deconvolution method of Lucy (1974). 
The iteration process was halted at the point when
\begin{equation}
\sqrt{\frac{(\Sigma_{\rm HI}-\Sigma_{\rm model})^{2}}{\Sigma_{\rm HI}^{2}}} < 10^{-5}\ ,
\end{equation}
where $\Sigma_{\rm HI}$ and $\Sigma_{\rm model}$ are the observed and modeled H{\sc i} surface densities.
The profiles of the approaching and receding sides were averaged to yield a mean radial H{\sc i} profile
which is presented in the upper panel of Fig.~\ref{fig:deprojected_gas2683}.
The total H{\sc i} mass of the model is consistent with our observations.
As in most nearby spiral galaxies, the deprojected H{\sc i} surface density is about constant within the optical radius
(see, e.g., Leroy et al. 2008). Its value of $\sim~2$~M$_{\odot}$pc$^{-2}$ is remarkably low compared
to the typical value of $\sim 10$~M$_{\odot}$pc$^{-2}$ in nearby spiral galaxies: in the sample of Leroy et al. (2008),
only NGC~3351 and NGC~2841 show such low surface densities within the optical disk.
The inner surface density profile shows two peaks corresponding to an inner and outer ring structure
at radii of $3$~kpc and $8$~kpc. For larger radii the surface density decreases 
monotonically until $20$~kpc, before rising again toward a peak at $26$~kpc.
We interpret and model the latter maximum as an H{\sc i} ring at the extremity of the gas disk.

The gas surface density is the sum of the atomic and molecular gas surface densities when assuming a Galactic CO-H$_{2}$ conversion factor
(upper panel of Fig.~\ref{fig:deprojected_gas2683}).
For the molecular gas surface density, we used the SEST CO(1-0) line flux of Elfhag et al. (1996)
and assumed an exponential distribution with a scale length of the stellar disk $40''=1.5$~kpc, 
as observed in local spiral galaxies (Leroy et al. 2008).
This led to a central surface density of molecular gas of  $40$~M$_{\odot}$pc$^{-2}$
(lower panel of Fig.~\ref{fig:deprojected_gas2683}). The atomic-to-molecular mass ratio of NGC~2683 is about 5.
This ratio is located at the lower end of the atomic-to-molecular mass ratio distribution for Sb galaxies (Fig.~4 of Young \& Scoville 1991).

For the mass modeling of NGC~2683, we applied GALFIT (Peng et al. 2002) to the 2MASS K' image
to obtain surface brightness profiles for the bulge and disk components of NGC~2683.
The derived parameters (central surface density $\Sigma_{0}$, surface density at the effective radius of the Sersic 
profile\footnote{The Sersic profile is given by $\Sigma=\Sigma_{\rm e} \exp\big(-\kappa ((R/R_{\rm e})^{1/n}-1)\big)$
with $\kappa=2n-0.331$.}
$\Sigma_{\rm e}$, scale length $l$, effective radius $R_{\rm e}$ of the Sersic profile,
Sersic index $n$, and total stellar mass M$_{*}$ assuming a K' band mass-to-light ratio of $1.0$) are given in Table~\ref{tab:params}.
\begin{table*}
      \caption{GALFIT derived model parameters of NGC~2683}
         \label{tab:params}
         \begin{tabular}{llllllc}
           \hline
            & $M/L_{\rm K'}$ & $\Sigma_{\rm e}/\Sigma_{0}$ & $R_{\rm e}/l$ & $n$ & $M_{*}$ & axis ratio \\
            & & (M$_{\odot}$pc$^{-2}$) & (kpc) & & ($10^{10}$ M$_{\odot}$) & \\
           \hline
           Bulge & $1.4$ & $184 \pm 4$ & $1.68 \pm 0.07$ & $5.20 \pm 0.09$ & $1.20 \pm 0.02$ & 0.7 \\
           Disk & $0.9$ & $1825 \pm 43$ & $1.50 \pm 0.01$ & $1.0$ & $2.58 \pm 0.05$ & - \\
           \hline
         \end{tabular}
\end{table*}
The K' band stellar disk scale length is $l_{*}=1.5$~kpc, a
low value for a rotation velocity similar to that of the Galaxy or galaxies of comparable rotation velocities 
(Fig.~8 of Courteau et al. 2007).
The decomposition of the rotation curve for different K' band mass-to-light ratios
from $0.5$ to $1.3$ is shown in the lower panel of Fig.~\ref{fig:deprojected_gas2683}. 
Whereas the fit of the exponential disk is robust, the fit of the bulge component depends 
somewhat on the initial conditions that were used in GALFIT. After inspecting the residual image, we are confident
that the mass profile is reproduced within the given errors (Table~\ref{tab:params}).

We assumed that the dark matter halo is isothermal with a core following the density distribution:
\begin{equation}
\rho(R)=\frac{\rho_{0}}{1+(R/R_0)^2}\ ,
\end{equation}
where $\rho_0$ is the central density and $R_0$ the core radius. 
Alternatively, we used an NFW profile (Navarro et al. 1996):
\begin{equation}
\rho(R)=\frac{\rho_0}{\frac{R}{R_{\rm S}}(1+\frac{R}{R_{\rm S}})^2}\ ,
\end{equation}
where $\rho_0$ is the central density and $R_{\rm S}$ the scale radius.
For a given K' band mass-to-light ratio M/L$_{\rm K'}$ for the bulge and the disk, the halo central density
and core radius were determined by minimizing $\chi^2$ between the model and observed rotation curves.
We calculated $\chi^2$ for a grid of (M/L$_{\rm K'})_{\rm bulge}$, (M/L)$_{\rm disk}$, $\rho_0$, and $R_0$
yielding 
\begin{itemize}
\item
for the cored isothermal halo ($\chi^2=3.10$):
(M/L$_{\rm K'})_{\rm bulge}$=1.4, (M/L$_{\rm K'})_{\rm disk}$=0.9, $\rho_0=0.16$~M$_{\odot}$\,pc$^{-3}$ and $R_{0}=2.0$~kpc; and
\item
for a NFW halo ($\chi^2=3.63$):
(M/L$_{\rm K'})_{\rm bulge}$=1.2, (M/L$_{\rm K'})_{\rm disk}$=0.9, $\rho_0=0.03$~M$_{\odot}$\,pc$^{-3}$ and $R_{0}=7.5$~kpc.
\end{itemize}
based on the $\chi^2$ values, the cored isothermal halo is the preferred model.

For our modeling, we adopted a mass-to-light ratio of $M/L_{\rm K'}=1.4$ for the bulge and $M/L_{\rm K'}=0.9$ for the disk component.
The latter value is compatible with the value of $0.8$ given by Bell et al. (2003; appendix~\ref{app}).
However, it is situated at the very high end of the $M/L_{\rm K'}$ distribution determined by the
DiskMass Survey (Fig.~1 of Martinsson et al. 2013).
Lower values of the mass-to-light ratio lead to gas velocity dispersions lower than typically
observed in the inner disks of nearby spiral galaxies (Tamburro et al. 2009; see Sect.~\ref{sec:vvd}).

The GALFIT absolute magnitude of $M_{\rm K_{\rm s}}=6.328 \pm 0.017$ is consistent with the value of $M_{\rm K_{\rm s}}=6.30 \pm 0.01$
from the 2MASX catalog (Skrutskie et al. 2006).
The disk/bulge decomposition gave disk and bulge masses of $2.6 \times 10^{10}$~M$_{\odot}$ and
$1.2 \times 10^{10}$~M$_{\odot}$, respectively.
We thus determined a total stellar mass of $M_{*}=3.8 \times 10^{10}$~M$_{\odot}$.
This is quite low for a rotation velocity of $220$~km\,s$^{-1}$ (Fig.~8 of Courteau et al. 2007). 

To investigate the 3D structure of the atomic gas disk, we made different 3D axisymmetric models for which 
we produced model H{\sc i} data cubes. These models have the following components:
\begin{itemize}
\item
a thin gas disk with a thickness of $500$~pc (TD), 
\item
a radially decreasing velocity dispersion (Vd),
\item
an asymmetric disk structure (E),
\item
a possible warp in inclination and position angle of the thin disk (PA, INC),
\item
different gas flares at galactic radii larger than $9$~kpc (F0-F3; Fig.~\ref{fig:flare}),
\item
an outer gas ring ($R > 25$~kpc),
\item
a gas halo around the inner disk (H), and
\item
a vertically lagging rotation velocity (L).
\end{itemize}

For the inclination angle of the galactic disk, we set $90^{\circ} \leq i \leq 60^{\circ}$.
The comparison of the model position velocity diagrams with those of the high-resolution C+D array data 
showed that $i=83^{\circ}$ is the best choice for the inner high surface density disk. 
In addition, it turned out that $i=87^{\circ}$ for the outer H{\sc i} ring ($R > 24$~kpc) led to the best resemblance 
between the model and observed gas distributions (column TD in Fig.~\ref{fig:PV1HR}). 
For galactic radii between $10$ and $24$~kpc, we considered warps in inclination and position angle.

To reproduce the emission distribution in the D array channel maps close to the systemic velocity (Fig.~\ref{fig:canauxbestBR}),
we introduced a turbulent velocity dispersion that increases toward the galaxy center:
\begin{equation}
v_{\rm turb}= 10+10\, \exp(-(R/4~{\rm kpc})^{2})~{\rm km}\,{\rm s}^{-1}\ .
\end{equation}

For the flare, we assumed four characteristic radial profiles of the disk height $H=FWHM/2$ (Fig.~\ref{fig:flare}):
\begin{equation}
\rho (z) = \Sigma_0 (R)\, {\rm sech} (1.3\,z/H(R))\ ,
\label{eq:height}
\end{equation}
where $\Sigma_0 (R)$ is the surface density in the midplane.
We also tried an exponential and a ${\rm sech}^2$ (isothermal) dependence of the vertical density structure (see Sect.~\ref{sec:fflare}).

For radii larger than $R=10.5'=23.5$~kpc, which is the low surface density ring, the disk plane is shifted vertically
by $35''=1.3$~kpc. This was necessary because the two H{\sc i} emission blobs at the
extremities of the galactic gas disk are offset from the galaxy's major axis (see Sect.~\ref{sec:hidistribution}).

\subsection{Asymmetric structure of the disk \label{sec:asymms}}

Since the observed asymmetries in the position velocity diagrams (Sect.~\ref{sec:hikinematics}) cannot be reproduced
with an axisymmetric model, we modeled the inner gas disk with elliptical isodensity annuli. This choice is motivated by 
the existence of another non-axisymmetric structure in NGC~2683, a galactic bar within $R=2$~kpc$=54''$  
(Kuzio de Naray et al. 2009). Since the observed H{\sc i} asymmetries extend over more than $6'=13$~kpc in radius, 
a physical link between the inner bar structure and outer elliptical annuli is not obvious, though.

The size of the major axis of the elliptical component is $a=15$~kpc and the axis ratio is $b/a=2/3$. 
We varied the angle between the major axis and the plane of the sky
up to $90^{\circ}$ in steps of $10^{\circ}$. It turned out that an angle of $30^{\circ}$ reproduced our observations
in a satisfying way (column TD+E in Figs.~\ref{fig:PV1BR} and \ref{fig:PV1HR}).
This is similar to the result of Kuzio de Naray et al. (2009), who found the bar to be closer to side-on than end-on.
The approximate alignment between the two structures might occur by chance or indicate a common pattern speed of the bar 
and the outer non-axisymmetric gas structure, i.e. a dynamical coupling between the two structures.

\begin{figure}
        \resizebox{\hsize}{!}{\includegraphics{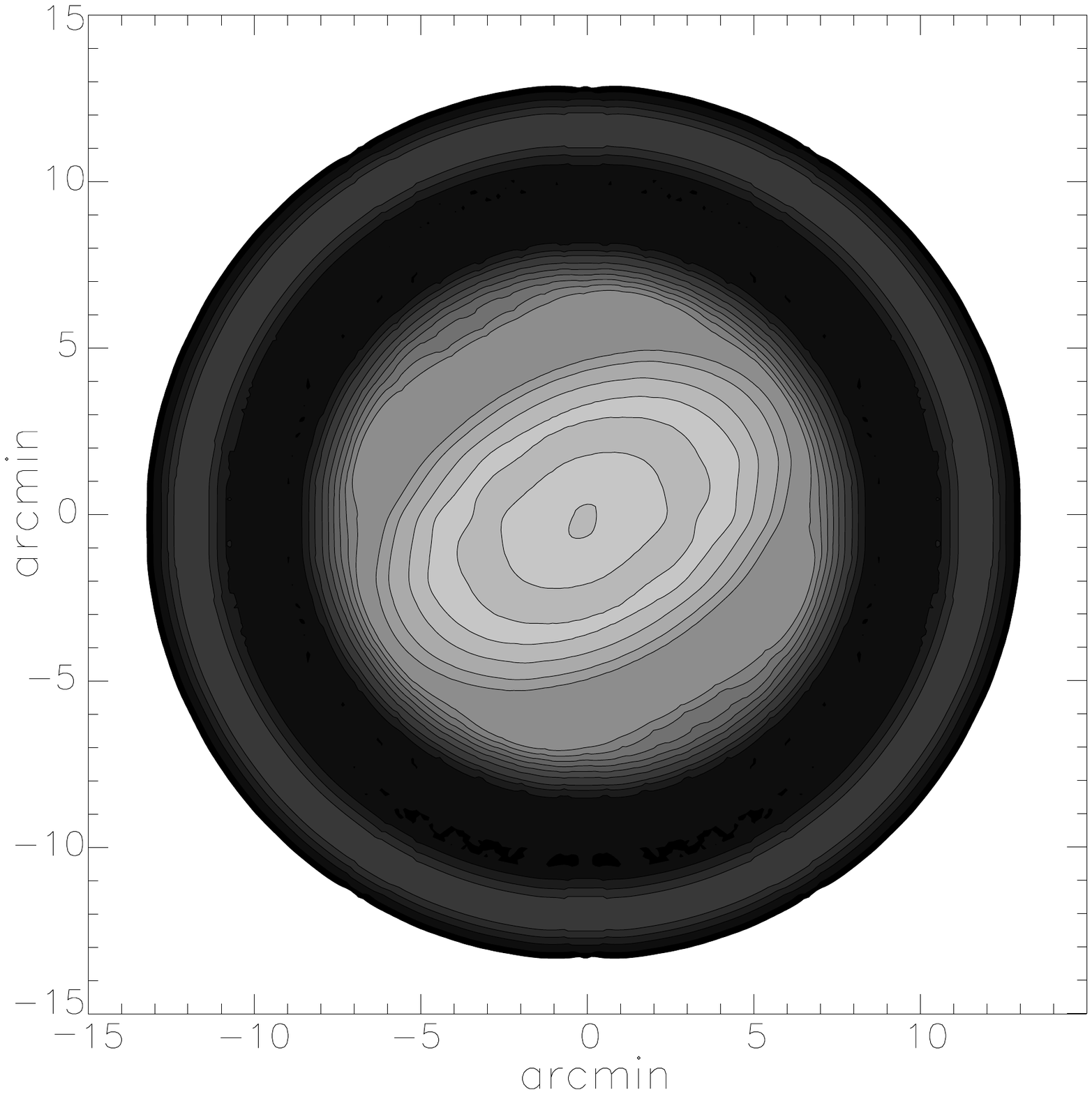}}
        \caption{Model gas distribution in face-on projection. 
          The resolution is $61'' \times 51''$. Grayscale levels are 
          $(1,2,4,6,8,12,16,20,24,28,32,48,64,96,128,192,264,392,784) \times 5$~mJy/beam\,km\,s$^{-1}$ or $2 \times 10^{18}$~cm$^{-2}$.
        }\label{fig:2683face}
\end{figure}

The resulting face-on projection of the gas distribution is presented in Fig.~\ref{fig:2683face}.
The resulting surface brightness profile (dash-dotted line in the upper panel of Fig.~\ref{fig:deprojected_gas2683}) 
does not differ significantly from the initially symmetric deprojected profile. 
The elliptical structure of the inner gas disk might be interpreted as a crude approximation to
spiral and/or bar structure (for more sophisticated models see Kamphuis et al. 2013).
This corresponds to an m=2 distortion that can be measured by a Fourier decomposition of the H{\sc i} distribution
(e.g., Bournaud et al. 2005). Compared to the THINGS sample (Walter et al. 2008), NGC~2683 resembles (i) the irregular galaxy NGC~4214, 
which shows an inner bar-like H{\sc i} structure  within a symmetric disk; (ii) the Sab galaxy NGC~4736, which shows an inner H{\sc i} ring with 
a different position angle than the main disk; and (iii) the Sc galaxy NGC~628, which shows an inner bar structure mainly in CO.
If the elliptical structure is mainly due to a bar, we expect a distorted velocity field, i.e. the 3D velocity vectors are approximately tangential
to the isophotes of the bar (P\'erez et al. 2004). If it is mainly due to spiral arms, we expect a much less distorted velocity field.

\begin{figure*}
        \resizebox{15cm}{!}{\includegraphics{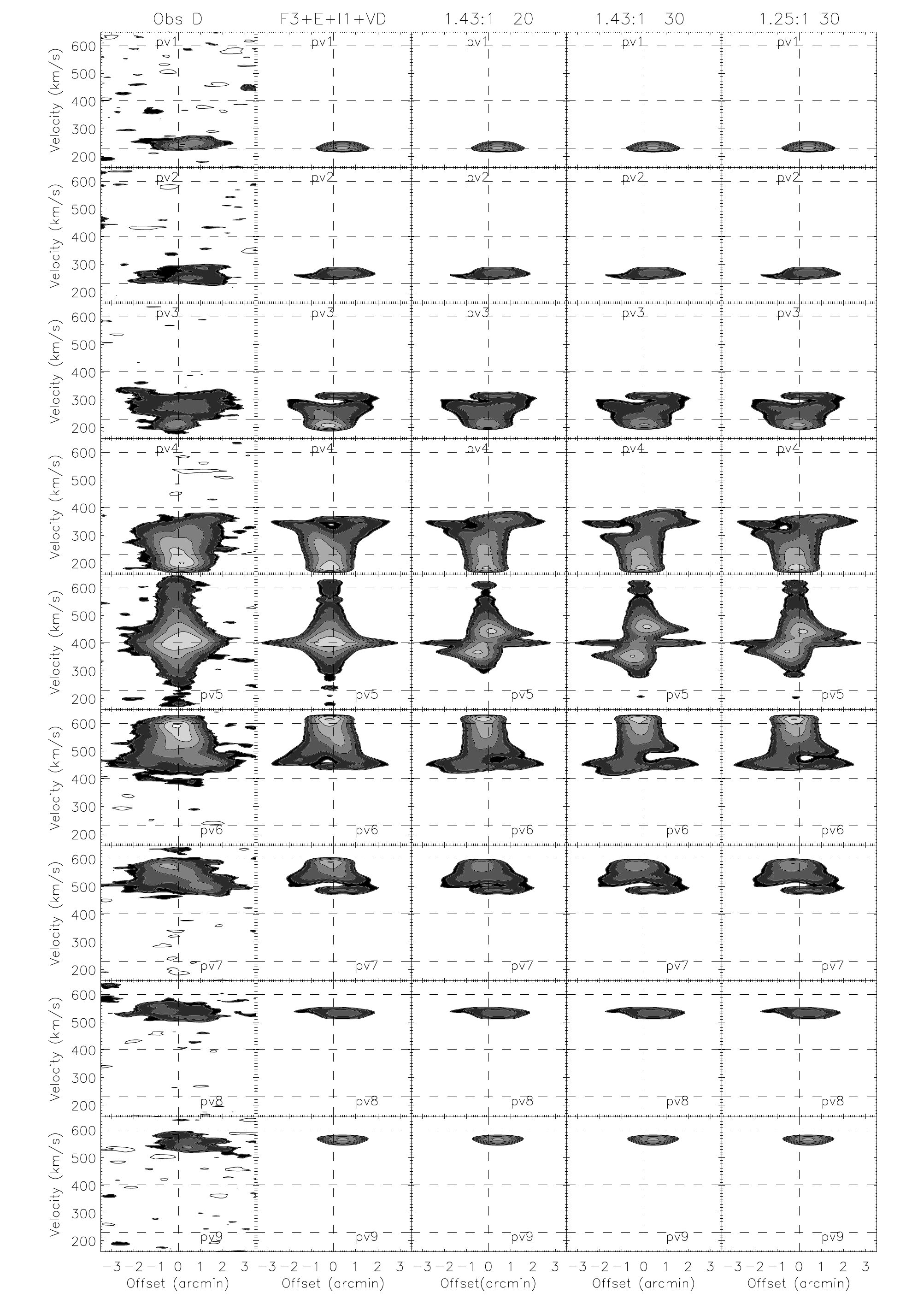}}
        \caption{NGC~2683 H{\sc i} D array and model position velocity diagrams.
          The contour levels are $(-2,2,3,6,12,24,48,96) \times 0.7$~mJy/beam.
          The resolution is $61'' \times 51''$. The model includes an elliptical component with axis ratios of $1.43:1/1.25:1$ and
          angles between the the bar's major axis and the plane of the sky of $20^{\circ}$ and $30^{\circ}$, a warp in inclination (I1),
          a centrally increasing velocity dispersion (VD), and a disk flare (F3).
        }\label{fig:PV_ellvel}
\end{figure*}

The influence of velocity distorsions on our model is shown in Fig.~\ref{fig:PV_ellvel} where the position velocity diagrams of a disk model with an elliptical
component and a velocity field following the bar isophotes is shown for different ellipticities and  different angles between the bar's major axis 
and the plane of the sky. While including an asymmetric velocity field improves the resemblance of pv3, it worsens
the resemblance of pv5, pv6, and pv7 to observations. Small velocity distortions are present in the inner disk of NGC~2683, because
we observe an asymmetric position velocity diagram in the center of NGC~2683 (pv 5). 
However, the observed amplitude of the asymmetry is much smaller than that of the models. 
We thus conclude that the velocity field of the elliptical component is rather symmetric.

\subsection{The disk warp \label{sec:warp}}

\begin{figure}
        \resizebox{\hsize}{!}{\includegraphics{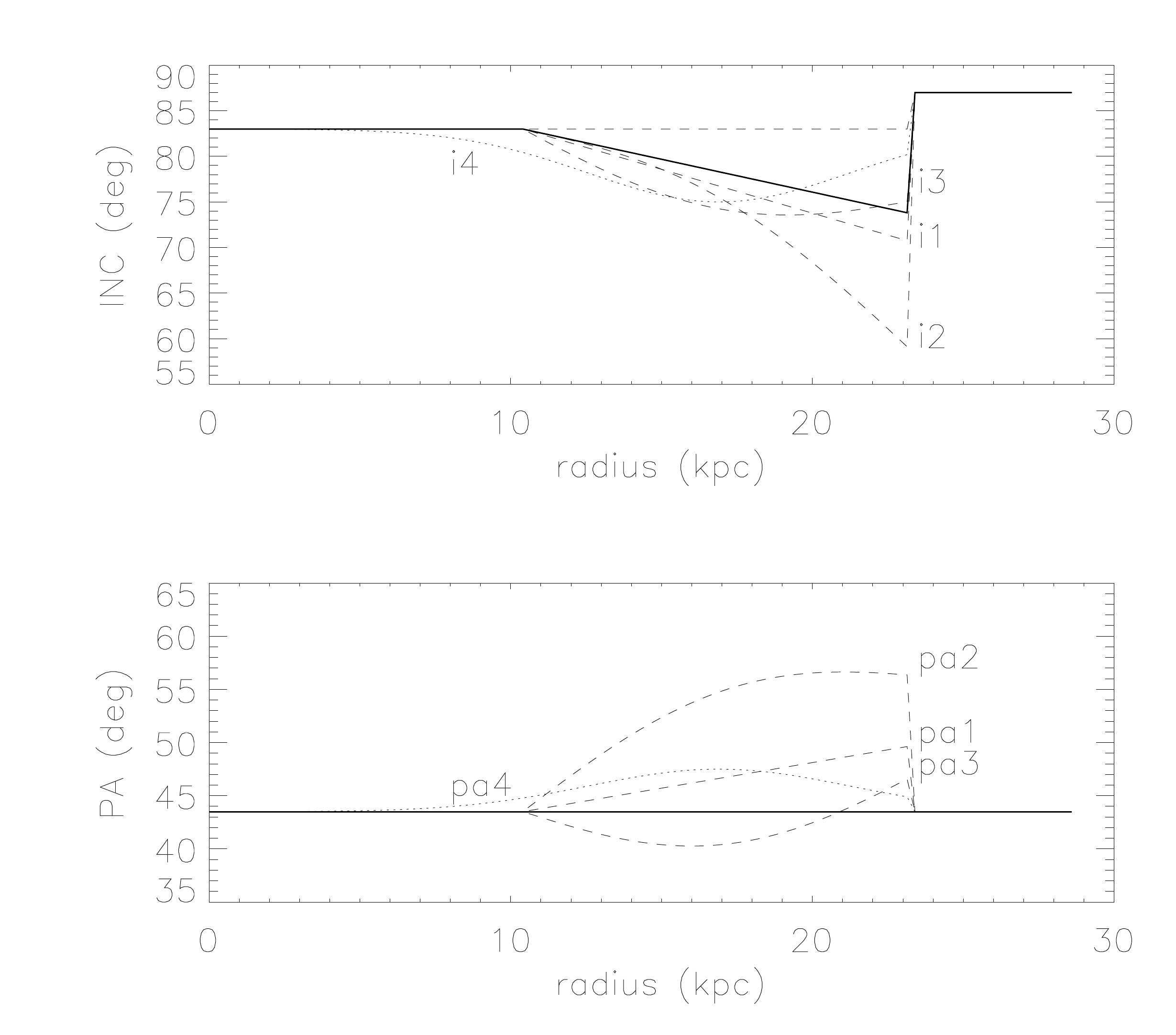}}
        \caption{Model inclination and position angle with respect to the galactic radius. The best-fit
          parameters are shown with a solid line. The dashed and dotted lines represent the tested variations for
          discarded models. The model with profiles represented by dotted lines are not shown in Fig.~\ref{fig:PV-10-4-TD}.
        }\label{fig:incli_pa}
\end{figure}

\begin{figure*}
        \resizebox{15cm}{!}{\includegraphics{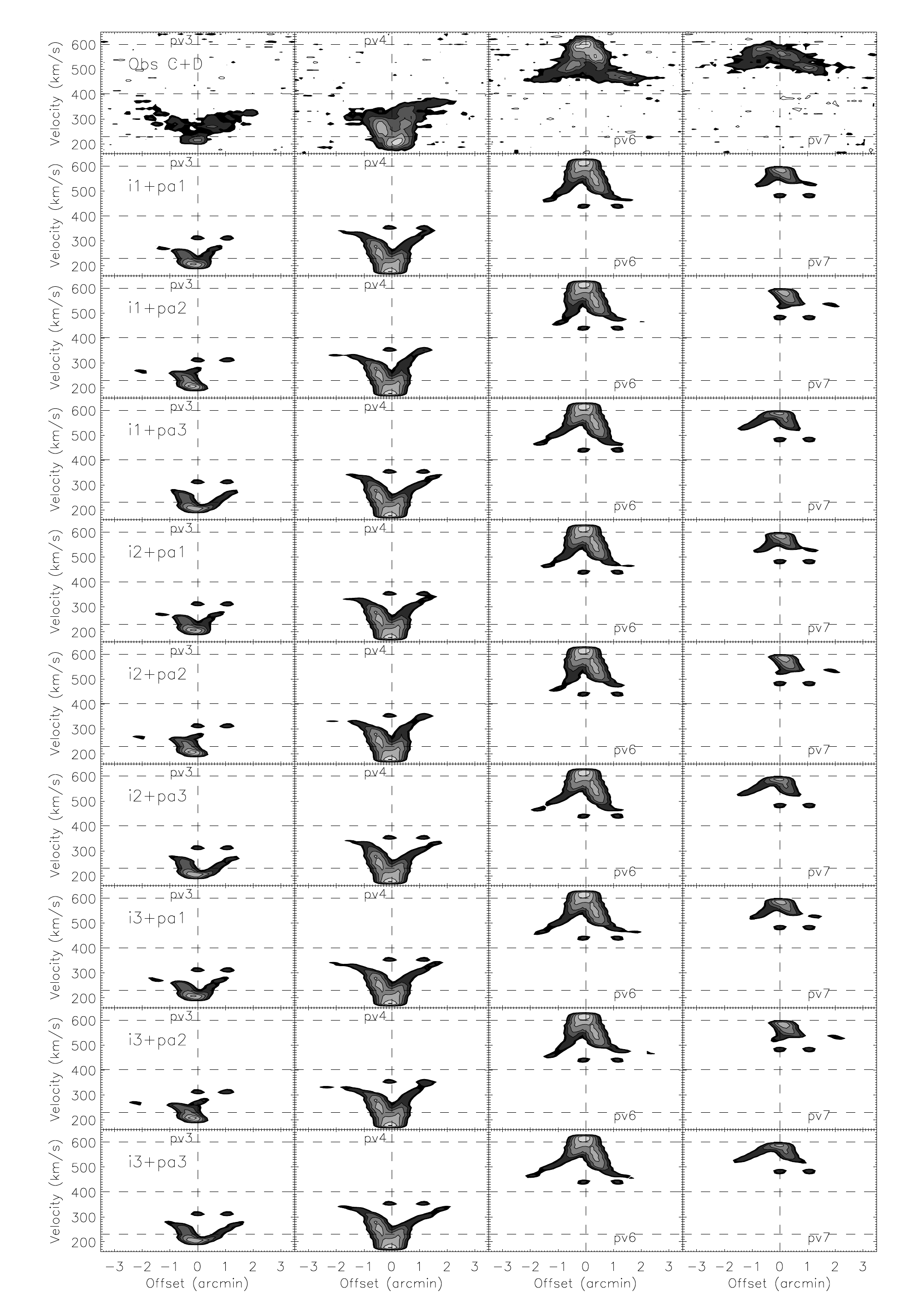}}
        \caption{NGC~2683 H{\sc i} C+D array and model position velocity diagrams.
          The contour levels are $(-2,2,3,6,12,24,48,96) \times 0.3$~mJy/beam.
          The resolution is $19'' \times 18''$. The different warp models refer to the profiles of Fig.~\ref{fig:incli_pa}.
        }\label{fig:PV-10-4-TD}
\end{figure*}

Can the projected thickness of the H{\sc i} disk be caused by a warp of the thin disk? 
To answer this question, we chose five characteristic inclination and four position angle profiles to create tilted ring
models of the thin H{\sc i} disk (Fig.~\ref{fig:incli_pa}). 
The radial extent of the warp is set by the fact that warps typically begin at an H{\sc i} column density of 
$\sim 2$~M$_{\odot}$pc$^{-2}$ where the H{\sc i} density drops down (Garcia-Ruiz et al. 2002).
The maximum inclination/position angles of the model profiles were determined by the observed projected disk thickness.
We created model cubes for all combinations of the inclination and position angle profiles.
For clarity, we only show selected C+D array position velocity diagrams of nine of these combinations in Fig.~\ref{fig:PV-10-4-TD}.

In the two position velocity diagrams of the approaching side (pv3 and pv4), the emission at negative
offsets is reproduced  well by the models. However, the linewidth of the emission at positive offsets is significantly
smaller than the observed linewidth. On the receding side (pv6 and pv7), we observe the
opposite trend: the  linewidth of the emission at negative offsets is significantly
smaller than the observed linewidth. In addition, the linewidth at offsets $> 1'$ are also smaller than observed.
These differences between our models and observations are present for all 20 models.
We thus conclude that the projected thickness of the H{\sc i} disk cannot be reproduced by a warp of the thin disk alone.

To reproduce the slight warp of at least a part of the gas disk between $7$~kpc and $14$~kpc observed in the
high and low resolution H{\sc i} distribution maps (Fig.~\ref{fig:ngc2683_d_mom0_1}), we selected a best-fit warp model
by eye based on the moment0 maps and position velocity diagrams (Fig.~\ref{fig:incli_pa}).
By only varying the position angle, we did not succeed in finding 
an axisymmetric model that reproduced the H{\sc i} gas distribution in a satisfying way.
Without a varying inclination angle, the spatial extent of the position velocity diagrams
pv3, pv4, pc6, and pv7 at velocities close to the systemic velocity is too small 
(Figs.~\ref{fig:PV2BR} and \ref{fig:PV2HR}).
Varying the inclination by $10^{\circ}$ toward lower inclination angles improved the observed position velocity diagrams in a satisfying way
(column TD+E compared to TD+E+INC of Figs.~\ref{fig:PV1BR} and \ref{fig:PV1HR}). However, the
spatial extent of the position velocity diagrams, especially beyond the optical disk (pv1, pv2, pv3, pv7, pv8, pv9),
are significantly smaller than our observations.

Thus, the H{\sc i} data is consistent with a thin disk component of constant position angle and a change in inclination by 
$10^{\circ}$ between $10$~kpc and $23$~kpc. A second thick component is needed beyond the optical disk.
The outer low surface brightness ring is modeled best with an inclination of $87^{\circ}$.

\subsection{The disk flare \label{sec:fflare}}

\begin{figure}
        \resizebox{\hsize}{!}{\includegraphics{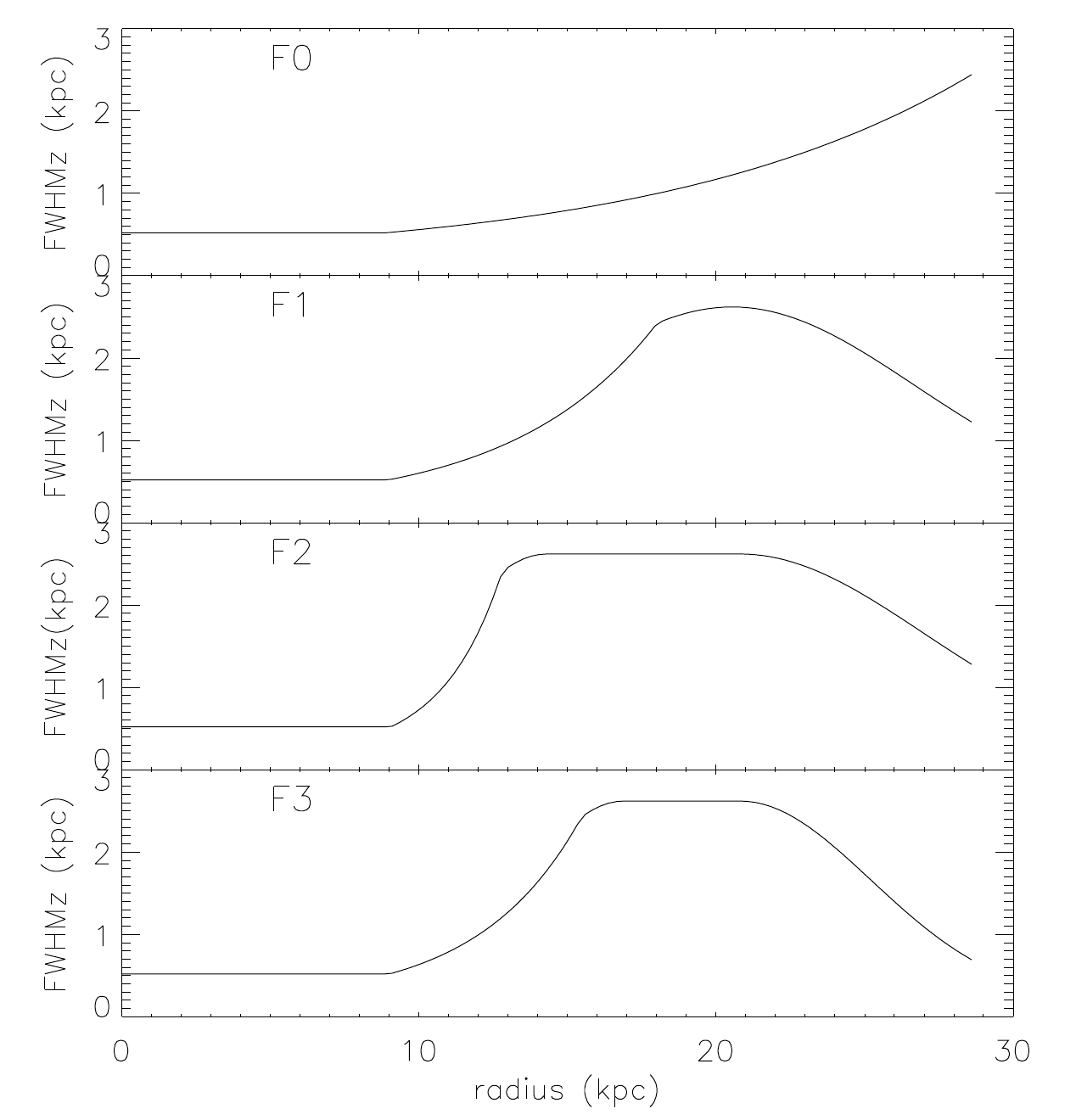}}
        \caption{Model disk flares.
        } \label{fig:flare}
\end{figure}

As shown in the previous section, the observed position velocity diagrams beyond the optical radius 
cannot be reproduced by a tilted ring model that only 
includes a thin disk (columns TD in Figs.~\ref{fig:PV1BR} and \ref{fig:PV1HR}). We therefore introduced a flaring gas disk into our model.
The disk flare was modeled by an exponential rise in the disk height $H(R)$=FWHM$/2$. 
The maximum disk width FWHM is set by the comparison between (i) the outer contours of the
model and observed H{\sc i} moment 0 maps and (ii) the D array position velocity diagrams beyond the optical radius: 
FWHM=$2.6$~kpc. We varied the radial scale length of the
exponential rise of $H(r)$ between $9$~kpc and $2$~kpc. Moreover, we assumed that the flare saturates
at a width of $4$~kpc and might decrease at larger radii. The four different types of
warps that we tested in our model (F0 -- F3) are shown in Fig.~\ref{fig:flare}.

For the vertical distribution of the atomic gas density, we additionally tested an exponential and a ${\rm sech}^{2}$ profile 
(Fig.~\ref{fig:PV4BR}). We adapted the respective scale heights to obtain the same FWHM for all profiles. 
Whereas the ${\rm sech}^{2}$ profile is indistinguishable from the ${\rm sech}$ profile, the exponential vertical profile broadens 
the position velocity diagrams in the velocity direction. Since this broadening is consistent with most of the observed position
velocity diagrams, an exponential vertical distribution might be the better choice.
Keeping this in mind, we decided to keep the vertical ${\rm sech}$ profile of Eq.~\ref{eq:height} which is in between the isothermal
(${\rm sech}^2$) and the exponential case.

\subsection{Rotation velocity lag}

Thick H{\sc i} components with lagging rotation velocities are observed in NGC~891 (Oosterloo et al. 2007)
and NGC~3198 (Gentile et al. 2013), among others. While Oosterloo et al. (2007) modeled the thick component separately
by assigning it a lower rotation velocity,  Gentile et al. (2013) assumed a constant vertical gradient.
We assumed a linear vertical gradient of the following form:
\begin{equation}
v_{\rm rot} (z) = v_{\rm rot,0} - \xi \, (\frac{z}{1.0\ {\rm kpc}})\ ,
\end{equation}
where $v_{\rm rot,0}$ is the axisymmetric rotation velocity and $\xi=5\ ,10,\ 15~{\rm km\,s^{-1}}$.
These gradients are comparable to those found in NGC~891 and NGC~3198 ($\Delta v / \Delta z \sim 7-15$~km\,s$^{-1}$kpc$^{-1}$).
Since we are interested in a lagging H{\sc i} halo, different H{\sc i} halos with the different velocity lags
are presented in Fig.~\ref{fig:PV-HL-TD} (for the halo component, we refer to Sect.~\ref{sec:hihalo}).
The velocity lag makes the radial velocities of gas at high altitudes (or offsets) increase/decrease toward the
systemic velocity (see also Sect.~\ref{sec:hihalo}).

\section{Comparison with observations}

For the following comparison between models and observations, we started with an inclined thin disk model 
and then successively added a decreasing velocity dispersion (Vd), an elliptical component (E), a warp in position angle 
(PA) and inclination angle (INC), the different flare profiles (F0-F3), a rotation velocity lag (L), and an H{\sc i} halo (H).
For the models including a disk flare, only the best-fit warp of Sect.~\ref{sec:warp} was considered.
The gas distribution maps (Fig.~\ref{fig:mom0HR}) and nine position velocity diagrams along the galaxy's minor axis
were then calculated in the same
way as for the observations (Figs.~\ref{fig:PV2BR}, \ref{fig:PV2HR}, \ref{fig:canauxbestBR}, and \ref{fig:canauxbestHR}).

\subsection{Position velocity diagrams \label{sec:pvs}}

The addition of the elliptical disk component is necessary to reproduce the asymmetric emission distribution
in the position velocity diagrams of the inner gas disk (pv4, pv6, and pv7;  compare column TD with column TD+E of 
Figs.~\ref{fig:PV1BR} and \ref{fig:PV1HR}). 
The addition of a warp along the line of sight leads to an increased spatial extent of the V-shaped emission distribution 
(with two ``legs'') at velocities close to the systemic velocity in the position velocity diagram (the legs of the V are curved outwards). 
This leads to a closer resemblance between the model and observed position velocity diagrams of the inner disk mostly on the receding side 
(pv6 and pv7; compare column TD+E with column TD+E+i1 of Figs.~\ref{fig:PV1BR} and \ref{fig:PV1HR}).
On the approaching side, the warp model is consistent with the observed position velocity diagram pv3.
However, the increased spatial extent of the V-shaped emission distribution is not observed in pv4. 

\begin{figure*}
        \resizebox{15cm}{!}{\includegraphics{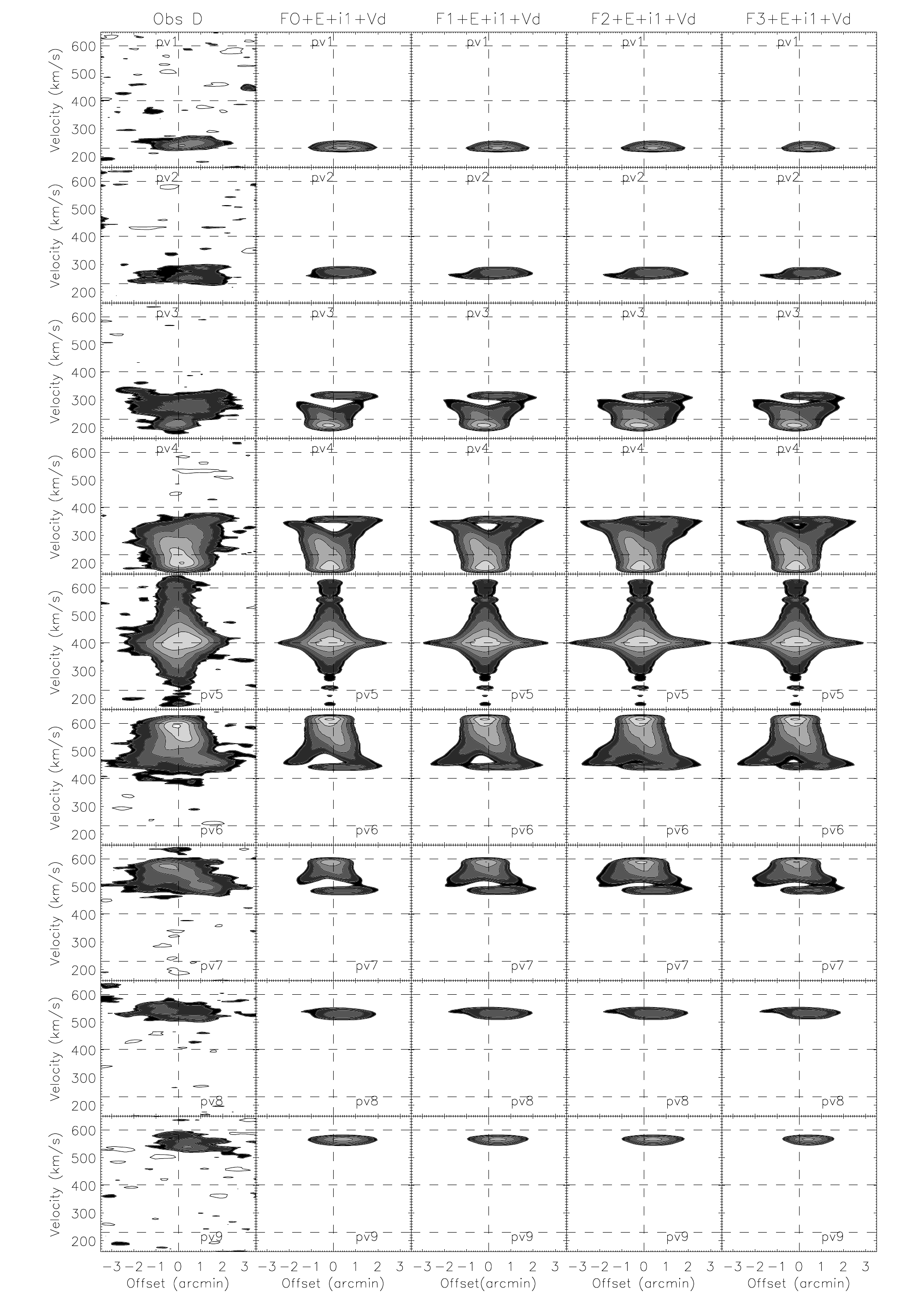}}
        \caption{NGC~2683 H{\sc i} D array and model position velocity diagrams.
          The model includes a thin disk, an elliptical component (E), a warp in inclination (i1), a radially decreasing velocity 
          dispersion (Vd), and a flare (F0-F3; see Fig.~\ref{fig:flare}).
          The contour levels are $(-2,2,3,6,12,24,48,96) \times 0.7$~mJy/beam.
          The resolution is $61'' \times 51''$.
        } \label{fig:PV2BR}
\end{figure*}

\begin{figure*}
        \resizebox{15cm}{!}{\includegraphics{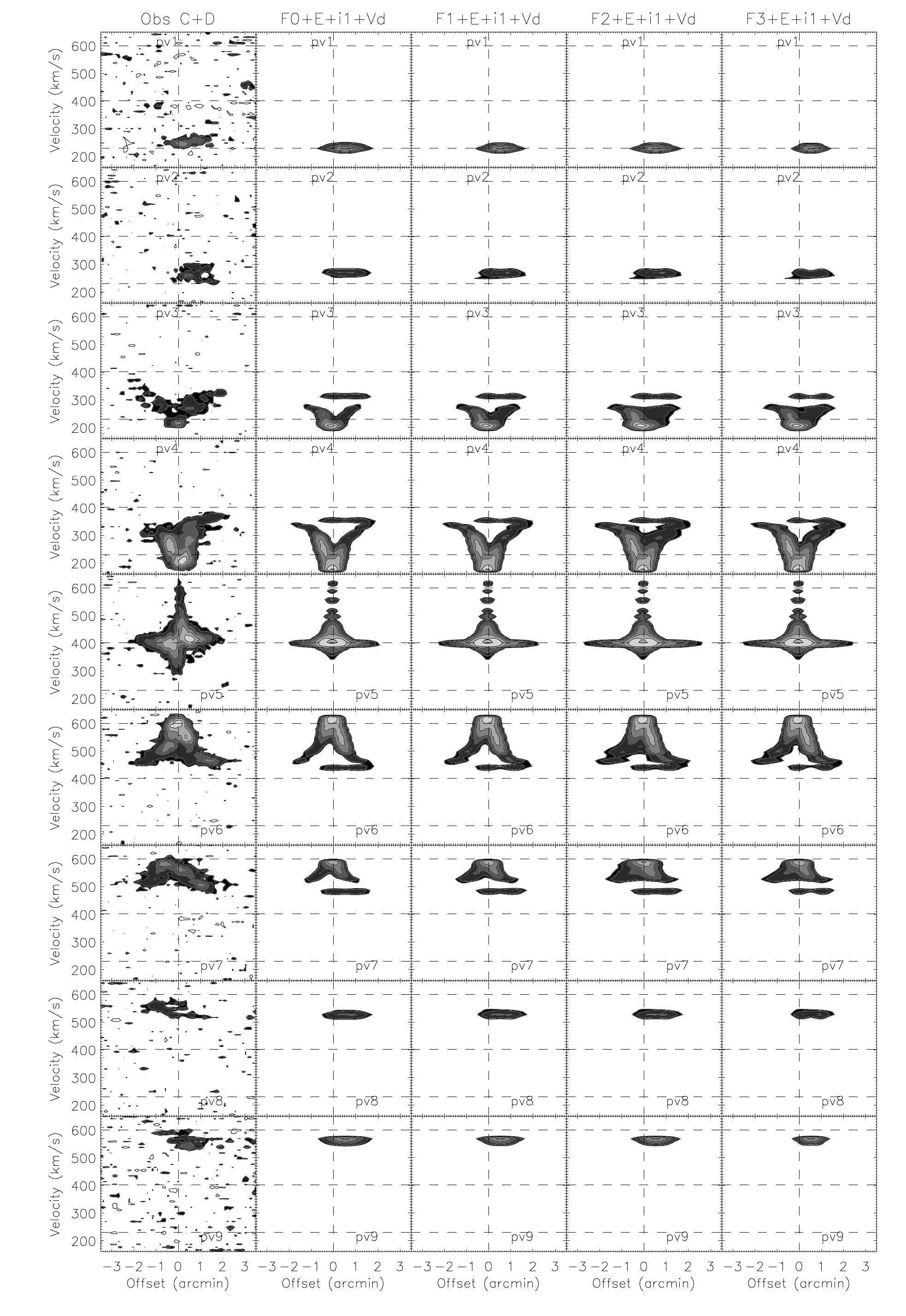}}
        \caption{NGC~2683 H{\sc i} C+D array and model position velocity diagrams.
          The model includes a thin disk, an elliptical component (E), a warp in inclination (i1), a radially decreasing velocity 
          dispersion (Vd), and a flare (F0-F3; see Fig.~\ref{fig:flare}).
          The contour levels are $(-2,2,3,6,12,24,48,96) \times 0.3$~mJy/beam.
          The resolution is $19'' \times 18''$.
        } \label{fig:PV2HR}
\end{figure*}

The different flare models lead to different spatial extents of the emission in the position velocity diagrams
at different galactic radii (flares F0 - F3 in Fig.~\ref{fig:PV2BR} (D array) and \ref{fig:PV2HR} (C+D array)).
We discuss these differences from negative (pv1) to positive offsets (pv9).
If not stated otherwise, the discussed features are present in the D and C+D array data.\\
pv1: Model F0/F3 shows the largest/smallest spatial extent of the H{\sc i} emission, comparable to or somewhat smaller than the observed extent. 
The model flares F0, F1, and F2 resemble the observed emission distribution most closely.\\
pv2: Models F0-F2 show the largest spatial extents of the high column density H{\sc i} emission, somewhat larger than the observed extent. 
The observed low column density extension to negative offsets (D array data) is not present in model F0. 
Model F3 resembles the observed emission distribution most closely.\\
pv3: The observed emission distribution is not reproduced well by the flare models. The legs of the V-shaped structure
are too thin for the model flares F0 and F1 (C+D array data). The spatial extent of the observed emission distribution is small
at low velocities, but significantly larger for the model F2 (C+D array data). The model F3 represents a compromise that
is marginally consistent with observations.\\
pv4: As for pv3, the legs of the V-shaped emission distribution are too thin for the model flares F0/F1 and too thick
for the model F2 in the C+D array data. This effect is less obvious in the D array data. 
The model F3 resembles the observed emission distribution most closely.\\
pv5: The observed asymmetric emission is not reproduced by our models.\\
pv6: As for pv3, the legs of the V-shaped emission distribution are too thin for the model flares F0/F1. The width of the
leg at negative offsets is reproduced better by model F2, and the leg at positive offsets by model F3. This is most visible in the C+D data.\\
pv7: The spatial extent of the observed emission distribution is small at low velocities and significantly 
larger for model F2. The legs of the V-shaped emission structures of the models flares F0 and F1 are thinner
than the corresponding observed structures (most visible in the C+D array data).
Model flare F3 represents a compromise that is most consistent with observations.\\
pv8: The observed high column density part of the emission distribution is located at somewhat higher velocities than the
low column density part. This feature is not reproduced by our models. The observed extent of the low column density emission
to negative offsets (D array data) is not present in model F0. Models F1-F3 approximately reproduce the observed emission distribution.\\ 
pv9: The spatial extent of the emission distribution of models F0, F1, and F2 are larger than observed.
Only model F3 reproduces the observed extent.

We conclude that model F3 is most consistent with our H{\sc i} observations.

\subsection{Selected channel maps}

\begin{figure*}
        \resizebox{\hsize}{!}{\includegraphics{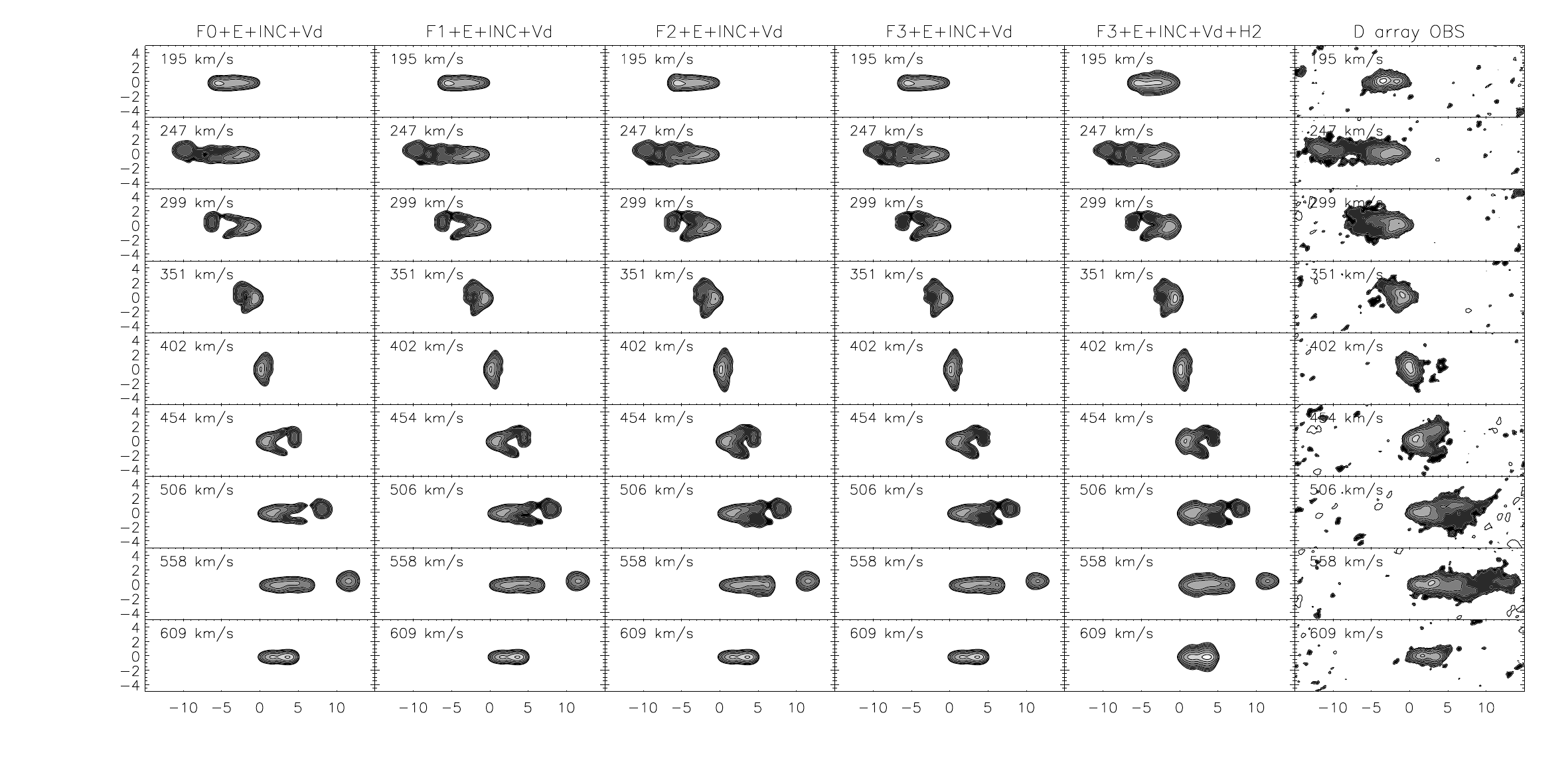}}
        \caption{Selected H{\sc i} D array and model channel maps.
          The model includes a thin disk, an elliptical component (E), a warp in inclination (INC), a radially decreasing velocity 
          dispersion (Vd), a flare (F0-F3; see Fig.~\ref{fig:flare}), and an H{\sc i} halo (H2).
          The offsets along the $x$- and $y$-axis are in arcmin.
          The contour levels are $(-2,2,3,6,12,24,48,96) \times 0.8$~mJy/beam.
          The resolution is $61'' \times 51''$. The galaxy was rotated by $45^{\circ}$.
        } \label{fig:canauxBR1}
\end{figure*}

\begin{figure*}
        \resizebox{\hsize}{!}{\includegraphics{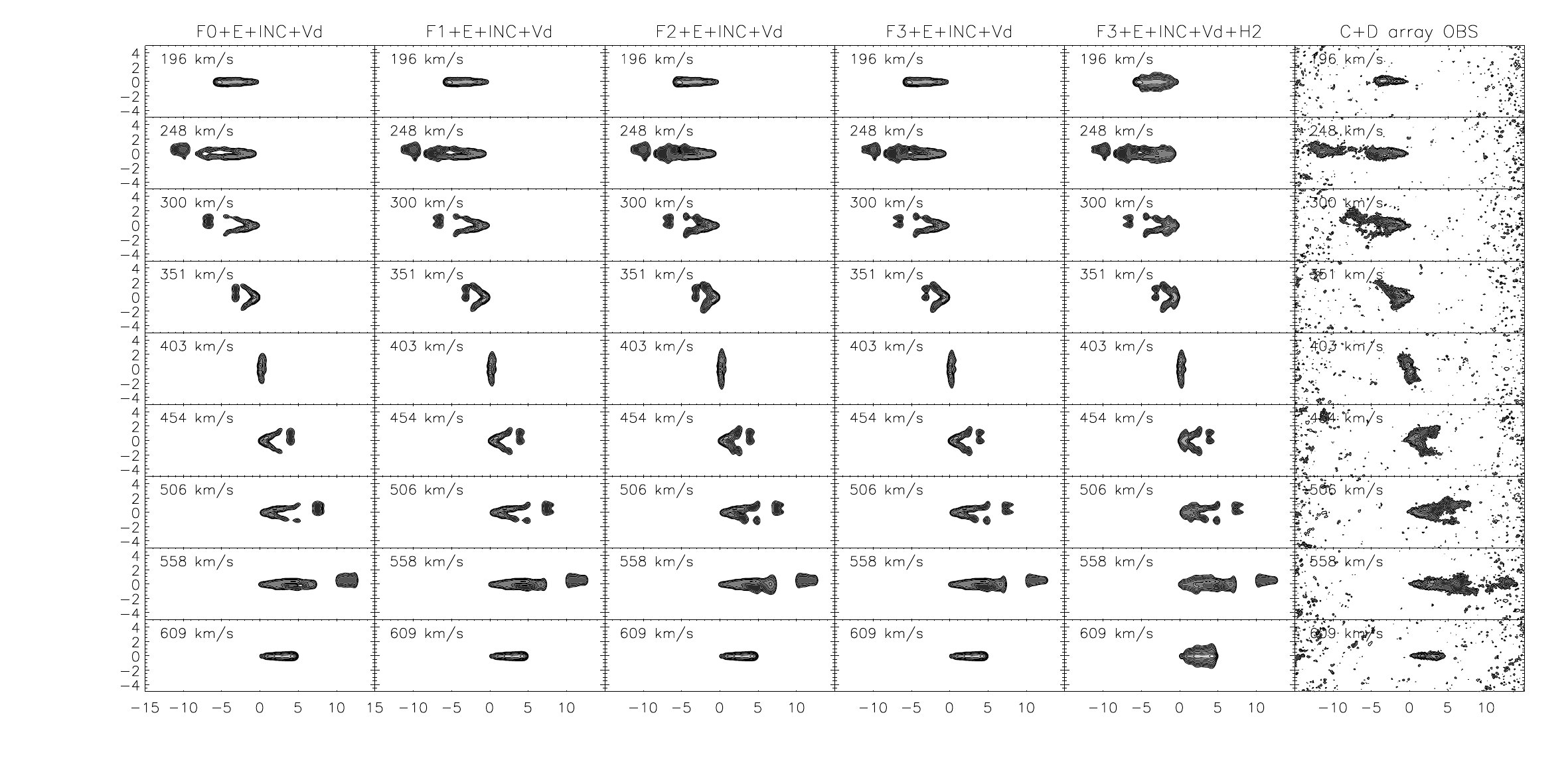}}
        \caption{Selected H{\sc i} C+D array and model channel maps.
          The model includes a thin disk, an elliptical component (E), a warp in inclination (INC), a radially decreasing velocity 
          dispersion (Vd), a flare (F0-F3; see Fig.~\ref{fig:flare}), and an H{\sc i} halo (H2).
          The offsets along the $x$- and $y$-axis are in arcmin.
          The contour levels are $(-2,2,3,6,12,24,48,96) \times 0.4$~mJy/beam.
          The resolution is $19'' \times 18''$. The galaxy was rotated by $45^{\circ}$.
        } \label{fig:canauxHR1}
\end{figure*}

Selected channel maps of the different models and the H{\sc i} observations are shown in Fig.~\ref{fig:canauxBR1} for the
D array and in Fig.~\ref{fig:canauxHR1} for the C+D array data.
A decreasing inclination angle leads to an increasing angle of the horizontal V-shaped
emission structure (the V becomes wider) in the channel maps. This is visible the most in the C+D data between $300$ and $506$~km\,s$^{-1}$
(Fig.~\ref{fig:canauxHR2}). The model with a varying inclination angle (TD+E+INC) reproduces observations
better than the model with a constant inclination angle (TD+E).

The addition of a flare to the model leads to an increased extent for the emission distribution in the vertical direction.
This effect is visible best in the channel maps between  $299$ and $508$~km\,s$^{-1}$ (Fig.~\ref{fig:canauxBR1}).
The legs of the horizontal V-shaped emission distribution are too thin for models F0 and F1 (at $299$, $454$, $508$~km\,s$^{-1}$).
The difference between models F2 and F3 is less obvious in the channel maps.

We conclude that models F2 and F3 reproduce the selected channel maps best.

\subsection{Atomic gas distribution \label{sec:atomic}}

\begin{figure*}
        \resizebox{16cm}{!}{\includegraphics{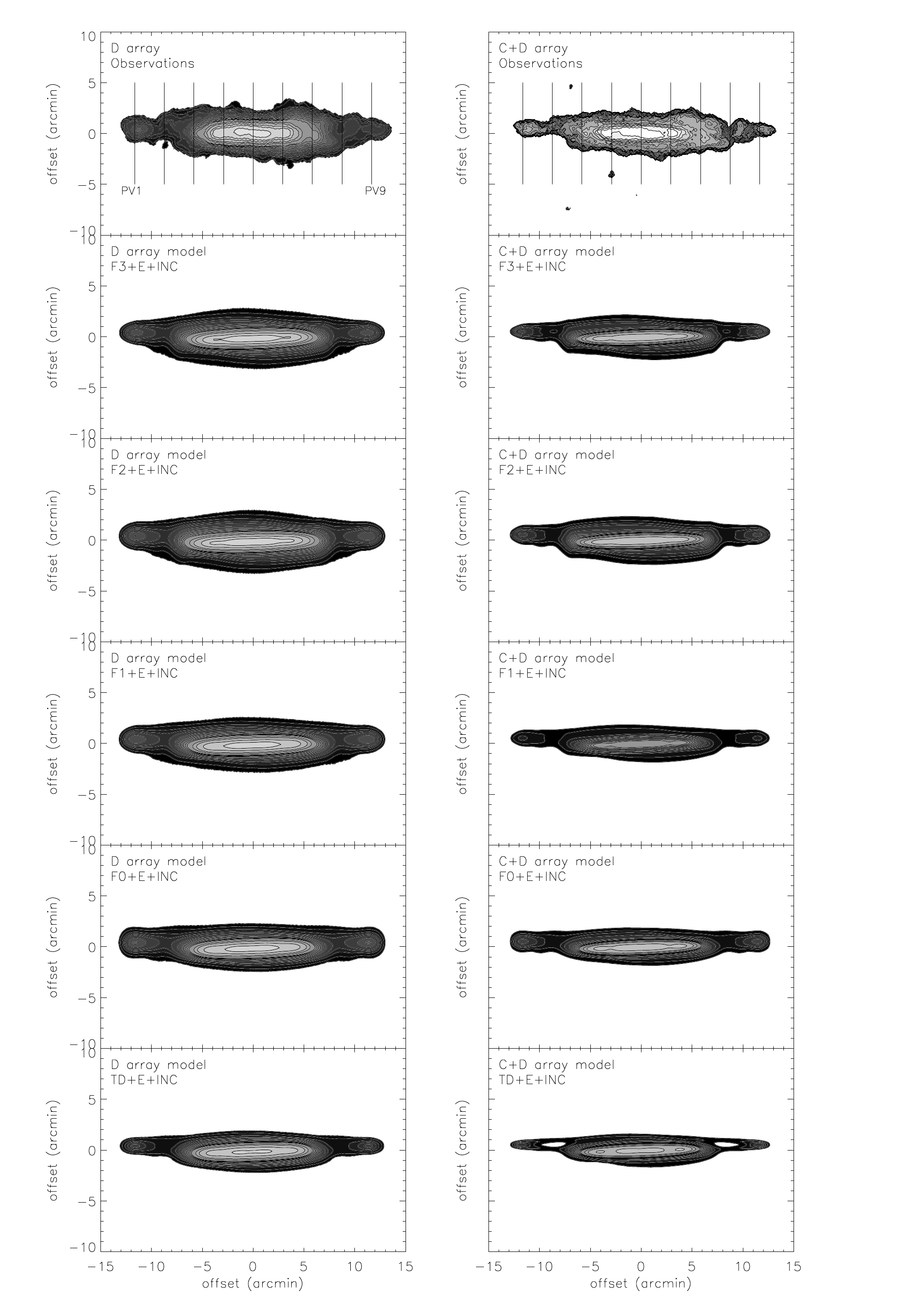}}
        \caption{Model gas distribution maps. Left column: D array; right column: C+D array data.
          Upper panels: observations; lower panels: models with an elliptical component (E), a warp in inclination (INC), and the warps
          F0-F3; best-fit (F3 + warp) model.
          The contours are the same as for Fig.~\ref{fig:ngc2683_d_mom0_1}.
        } \label{fig:mom0HR}
\end{figure*}

The atomic gas distributions of the models with different flares F0 to F3 are shown in Fig.~\ref{fig:mom0HR}
for the D and C+D array data.  
While the inner high column density H{\sc i} disk is barely affected by the different flare models,
the main differences are observed at the lowest column densities. Only models F2 and F3 show the observed
extent perpendicular to the major axis. We note that the largest vertical extent occurs along the minor axis
(owing to axisymmetry) in the models, whereas it occurs at an offset of $\sim 4'$ along the major axis in our observations.
This suggests that either the observed H{\sc i} flare is not axisymmetric or the disk volume
is not uniformly populated by atomic gas.
As observed in all model maps, two gas blobs, which we interpret as the lobes of a ring structure,
are located at the extremities of the gas disk ($R=11'=24.6$~kpc). 
By construction, they are vertically offset from the disks major axis.

We thus conclude that models F2 and F3 show the best resemblance to our H{\sc i} observations with
a slight preference for model F3 (Sect.~\ref{sec:pvs}).
In the following we refer to the model with the elliptical component, with an inclination varying by $10^{\circ}$, and flare F3 
as the best-fit model (F3+E+INC).

\begin{figure}
  \resizebox{\hsize}{!}{\includegraphics{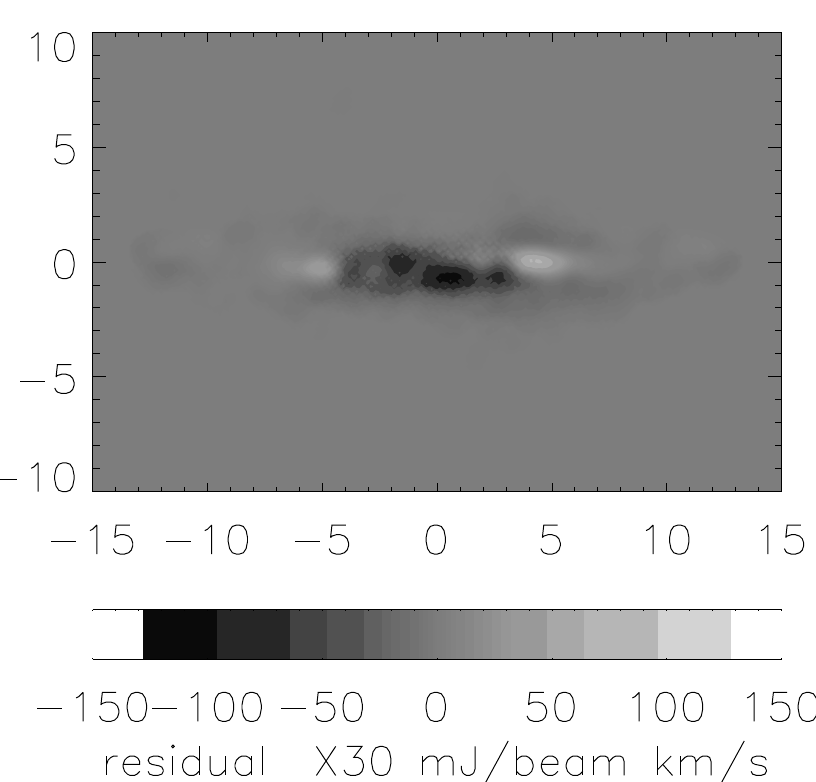}\includegraphics{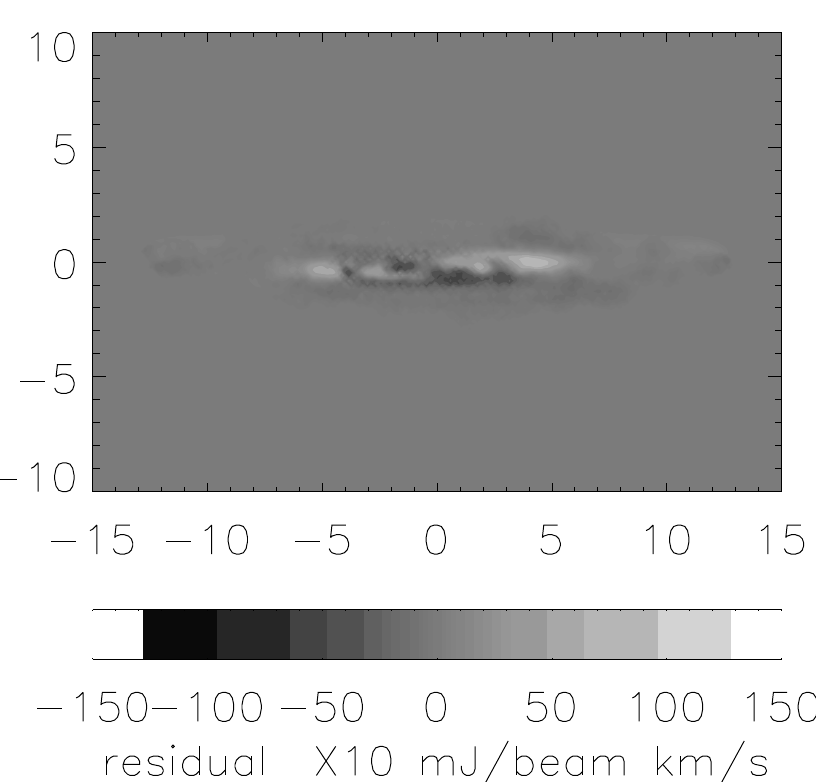}}
  \resizebox{\hsize}{!}{\includegraphics{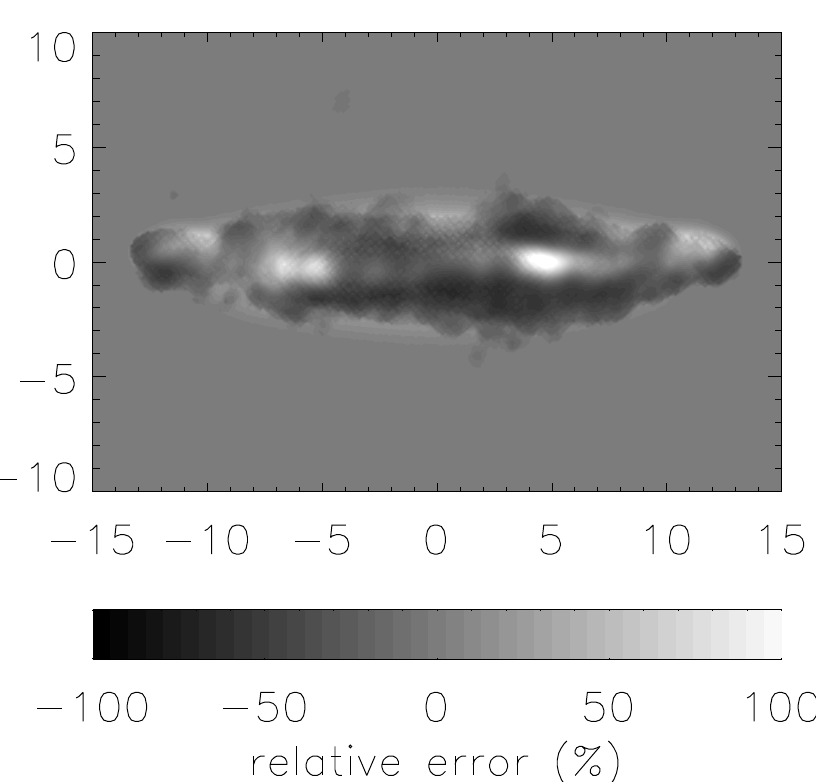}\includegraphics{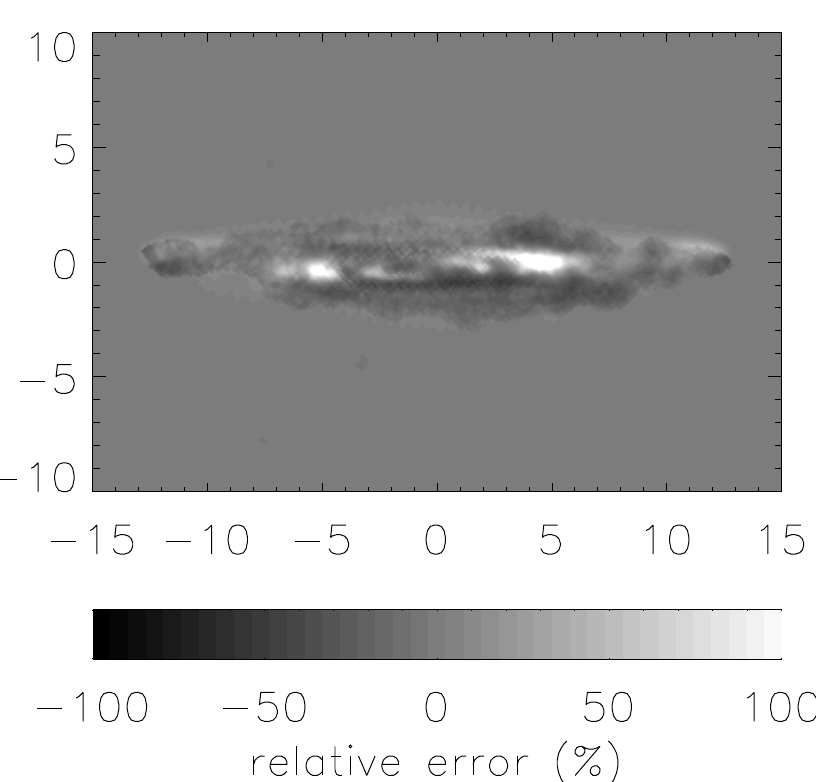}}
        \caption{Upper panels: residual column density from the best-fit model. Lower panels:
          relative residual column density from the best-fit model. Left: D array; right: C+D array data.
          The offsets along the $x$- and $y$-axes are in arcmin.
        } \label{fig:model_res}
\end{figure}

The relative and absolute residual column density of the best-fit model are presented in Fig.~\ref{fig:model_res}
for the D and C+D data separately.
Most residuals are asymmetric with respect to the major and minor axes and most probably caused by non-axisymmetric
features in the observed atomic gas surface density. 
Within the inner $4'$, negative residuals are located at positive/negative offsets on the $x$/$y$-axis 
and negative/positive offsets on the $x$/$y$-axis. These residuals are probably due to the spiral structure of the gas disk.
For galactic radii between $4'$ and $6'$, the residuals are mostly positive. 

\subsection{The rotation velocity lag}
 
Since the gas disk has a significant vertical extent only at galactic radii $>12$~kpc, the effects of 
the lag are only visible in pv1-pv3 and pv7-9, shifting the emission distribution toward velocities closer to the
systemic velocity. This only improves the resemblance between our model and observations for the outermost radius
(pv1 and pv9).
The addition of the rotation velocity lag changes the emission distribution significantly in four channels ($186$, $248$, $537$, and 
$599$~km\,s$^{-1}$), but does not significantly improve the resemblance between our models and observations.

We conclude that a rotation velocity lag is not a necessary ingredient for our model.

\subsection{Atomic gas halo \label{sec:hihalo}}

The origins of H{\sc i} halos are thought to be (i) a continuous galactic fountain flow, where gas is pushed up by stellar activity, 
travels through the halo and eventually falls back to the disc (Fraternali \& Binney 2006) or (ii) external gas accretion. 
For case (i) we expect an H{\sc i} halo within a radius
of $R < 10$~kpc where the molecular gas surface density and star formation activity is high (Fig.~\ref{fig:deprojected_gas2683} 
and \ref{fig:sfrspitzer}). For case (ii) the halo can extend to larger galactic radii. However, it can be most clearly
identified in the inner disk $R < 10$~kpc, where warps and flares are rare.

\begin{figure*}
        \resizebox{15cm}{!}{\includegraphics{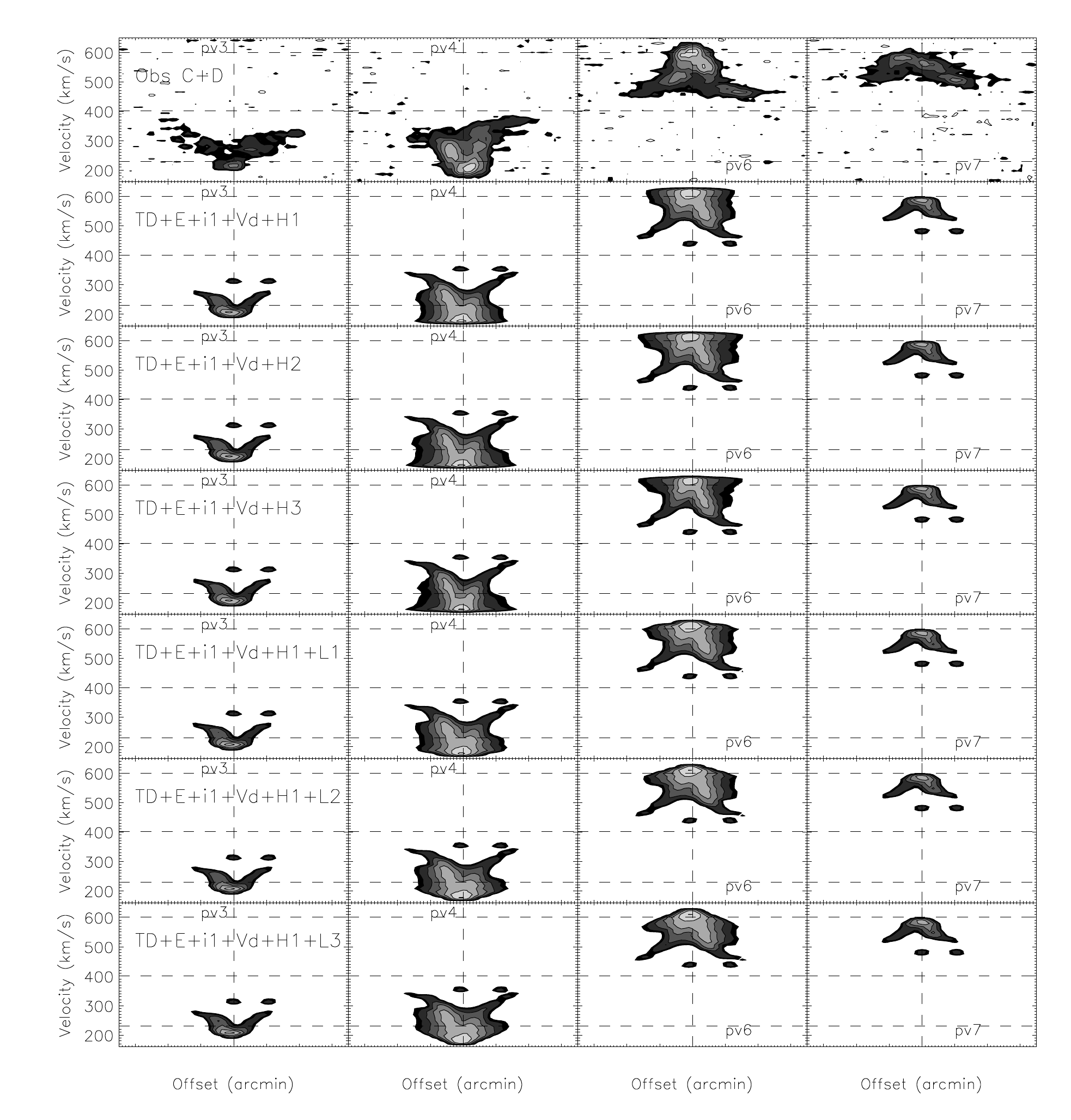}}
        \caption{Selected position velocity diagrams. C+D observations and models with a thin disk (TD), an elliptical
          component (E), a warp in inclination (i1), a centrally increasing velocity dispersion (Vd), different
          H{\sc i} halos (H1-H3), and different velocity lags (L1, L2, L3: $5,\ 10,\ 15$~km\,s$^{-1}$kpc$^{-1}$). 
          The contour levels are $(-2,2,3,6,12,24,48,96) \times 0.3$~mJy/beam.
          The resolution is $19'' \times 18''$.
        }\label{fig:PV-HL-TD}
\end{figure*}

Is there an extended atomic gas halo around the thin gas disk inside the optical radius?
To answer this question, we introduced an additional thick gas disk of the following form (Oosterloo et al. 2007):
\begin{equation}
\rho_{\rm halo}=\chi \, \rho_{\rm disk} \frac{{\rm sinh}(z/z_{0})}{{\rm cosh}(z/z_{0})^{2}}\ ,
\end{equation}
where $\rho_{\rm disk}$ is the density of the gas disk, and vertical $z_{0}$ the scale height in kpc.
We used the following combinations of $(\chi,z_{0})$: H1=$(0.3,0.75)$, H2=$(0.2,1.0)$, and H3=$(0.1,1.25)$. 
This thick gas halo does not show up in the position velocity diagrams nor in the selected channel maps of the low resolution 
(D array) model. 
However, it is clearly visible in the position velocity diagrams and channel maps of the inner disk (Fig.~\ref{fig:PV-HL-TD} 
pv3, pv4, pv6, and pv7) in the high resolution (C+D array) model in form of a low column density envelop around the thin disk.
The comparison with our preferred model (right panels of Fig.~\ref{fig:PV2HR}) shows that this additional envelop is not present in
the observed position velocity diagrams. 
This statement is confirmed by the direct comparison between the channel maps of the models with and without an H{\sc i} halo 
(Fig.~\ref{fig:canauxHR1}).

We conclude that NGC~2683 does not host an extended H{\sc i} halo around the optical and thin gas disk with column densities in excess 
of $\sim 3 \times 10^{19}$~cm$^{-2}$. The addition of a velocity lag does not change our conclusion.

\section{Discussion \label{sec:discussion}}

The main ingredients of our best-fit model of the gas disk are (i) a thin disk; (ii) an elliptical surface 
density distribution of the gas disk; (iii) a slight warp in inclination between $10$~kpc$ \leq R \leq 20$~kpc;
(iv) an exponential flare that rises from $0.5$~kpc at $R=9$~kpc 
to $4$~kpc at $R=15$~kpc, stays constant until $R=22$~kpc, and decreases its height for $R > 22$~kpc
(Fig.~\ref{fig:flare}); and (v) 
a low surface density gas ring with a vertical offset of $1.3$~kpc.
Based on the comparison between the high resolution model and observations, we exclude the possibility that there is an 
extended atomic gas halo around the optical and thin gas disk (Fig.~\ref{fig:PV-HL-TD}).

\subsection{Uncertainties of the derived parameters}

The absence of an objective measure for the goodness of our model fits led to it not being possible
to give precise uncertainties for the derived model parameters. Here we give estimates based on our experience with the comparison 
between our models and observations: the inclination angle of the thin disk is uncertain to $\Delta i \sim 3^{\circ}$.
The uncertainty of the rotation velocity $\Delta v_{\rm rot} \sim 10$~km\,s$^{-1}$ is mainly given by the spectral resolution. 
The lower limit of the K$'$ band mass-to-light ratio is determined by the typical velocity dispersion of nearby spiral galaxies.
We estimate the uncertainty to be $\Delta (M/L_{\rm K'}) \sim 0.3$. Thus, our preferred value of $M/L_{\rm K'}=0.9 \pm 0.3$
is significantly higher than the mean value of the DiskMass Survey ($M/L_{\rm K'} = 0.31 \pm 0.07$). Only one galaxy in this surveys
has a comparable mass-to-light ratio.
The uncertainty of the rotation curve of the dark matter halo  
is dominated by the uncertainty of the mass-to-light ratio in the inner disk ($\Delta v_{\rm DM} \sim 20$-$30$~km\,s$^{-1}$).
At $R > 10$~kpc, the latter uncertainty decreases and the uncertainty of the total rotation velocity is no longer negligible, yielding
$\Delta v_{\rm DM} \sim 15$~km\,s$^{-1}$.

For the disk warp (inclination and position angle), we used a limited number of characteristic
profiles. The best-fit profile of the inclination angle (Fig.~\ref{fig:incli_pa}) has a maximum variation of
$10^{\circ}$. We estimate the uncertainty on the radial variation of the profile to be on the order of $30$\,\%.
We did not observe any improvement in the model fit by a variation in the position angle. The uncertainty on this profile
is about $3$-4$^{\circ}$. For the disk flare, we also used a limited number (4) of characteristic profiles.
Since the model with flare F3 (Fig.~\ref{fig:flare}) reproduces our observations better (see Sect.~\ref{sec:pvs}), 
we suggest that the height of the flaring disk decreases at large galactic radii.
We can give a lower limit for the slope of the flaring part (model F1 is discarded) and a tendency to prefer a flare that does not rise 
too steeply (model F3 is somewhat preferred to model F1). Thus a flare profile between model profiles F2 and F3 is possible.
The maximum height of the flare is determined by the moment0 maps and the D array position velocity diagrams. 
It is uncertain to a fraction of the C+D array resolution, i.e. $\sim 0.3$~kpc. 

\subsection{Comparison with other H{\sc i} flares}

\begin{figure}
        \resizebox{\hsize}{!}{\includegraphics{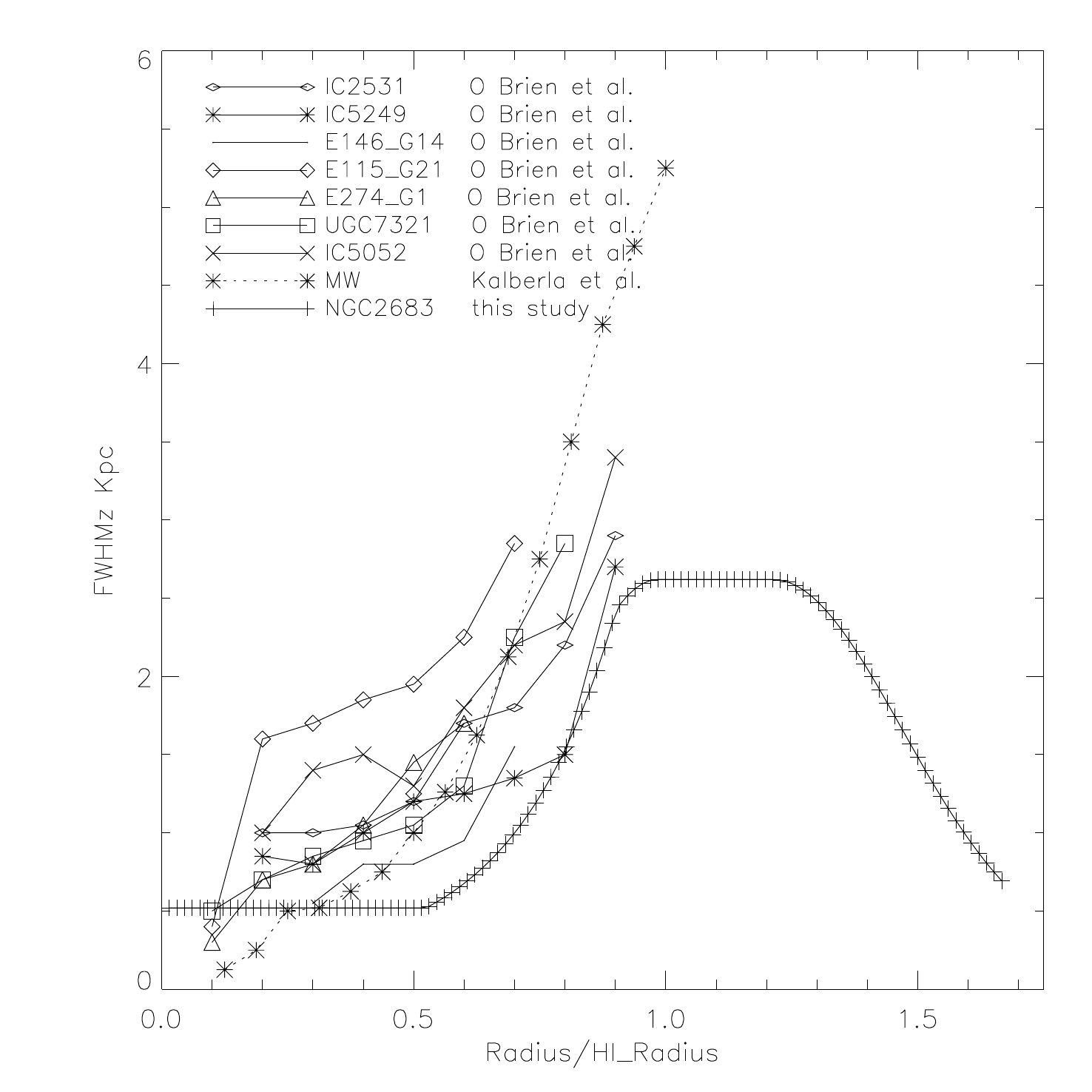}}
        \caption{Comparison between the flare of the gas disk and those of other spiral galaxies.
        } \label{fig:flare_OB}
\end{figure}

In Fig.~\ref{fig:flare_OB} we compare the flare of NGC~2683 with that of the Galaxy (Kalberla \& Kerp 2009) and
other nearby spiral galaxies (O'Brien et al. 2010). For this comparison we set the H{\sc i} radius of
NGC~2683 to $17$~kpc.
The slope of NGC~2683's flare is comparable, but somewhat steeper than those of the other spiral galaxies. 
NGC~2683's maximum height of the flare is also comparable with those of the other galaxies. On the other hand,
a saturation of the flare is only observed in NGC~2683.

\subsection{The vertical velocity dispersion of the gas \label{sec:vvd}}

\begin{figure}
        \resizebox{\hsize}{!}{\includegraphics{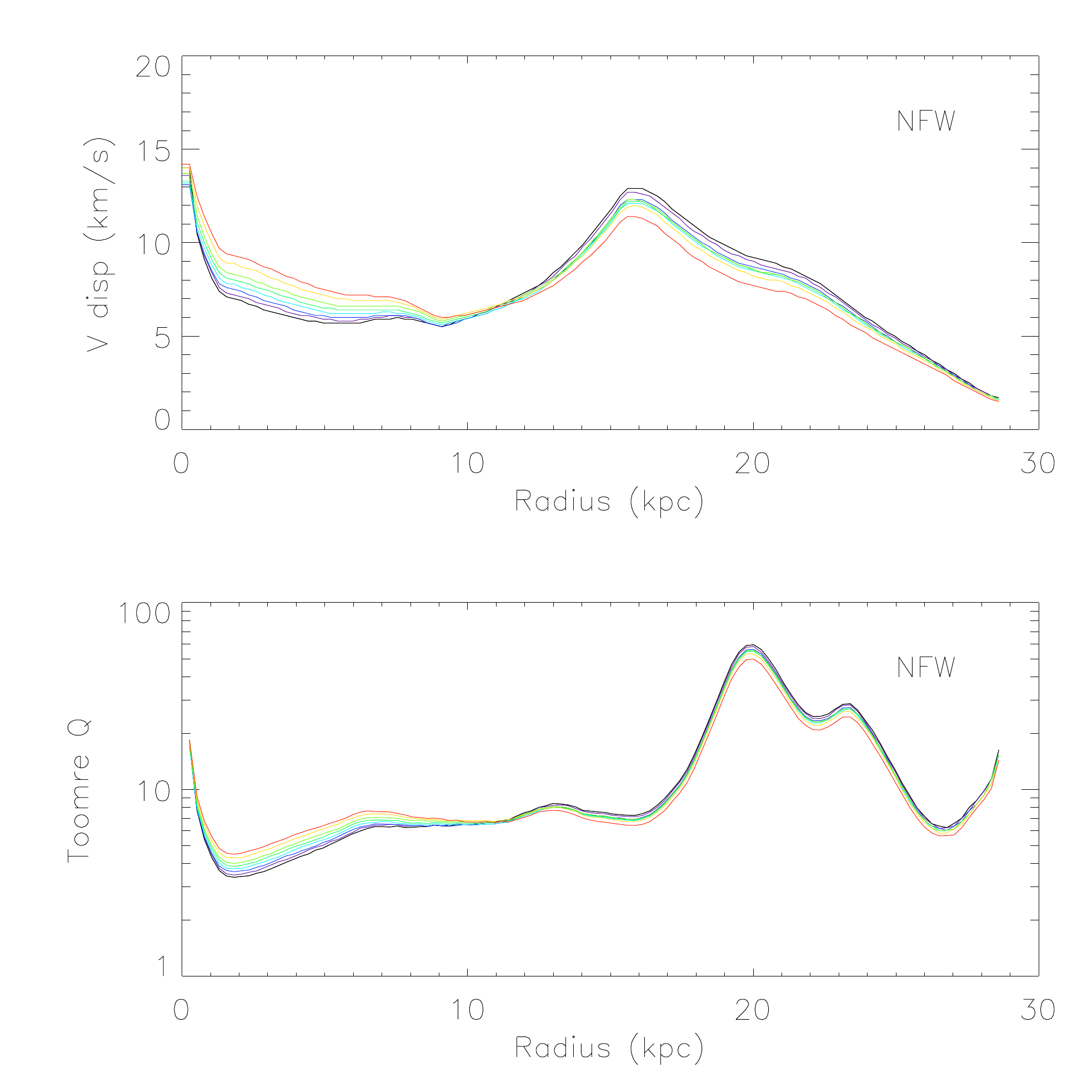}}
        \resizebox{\hsize}{!}{\includegraphics{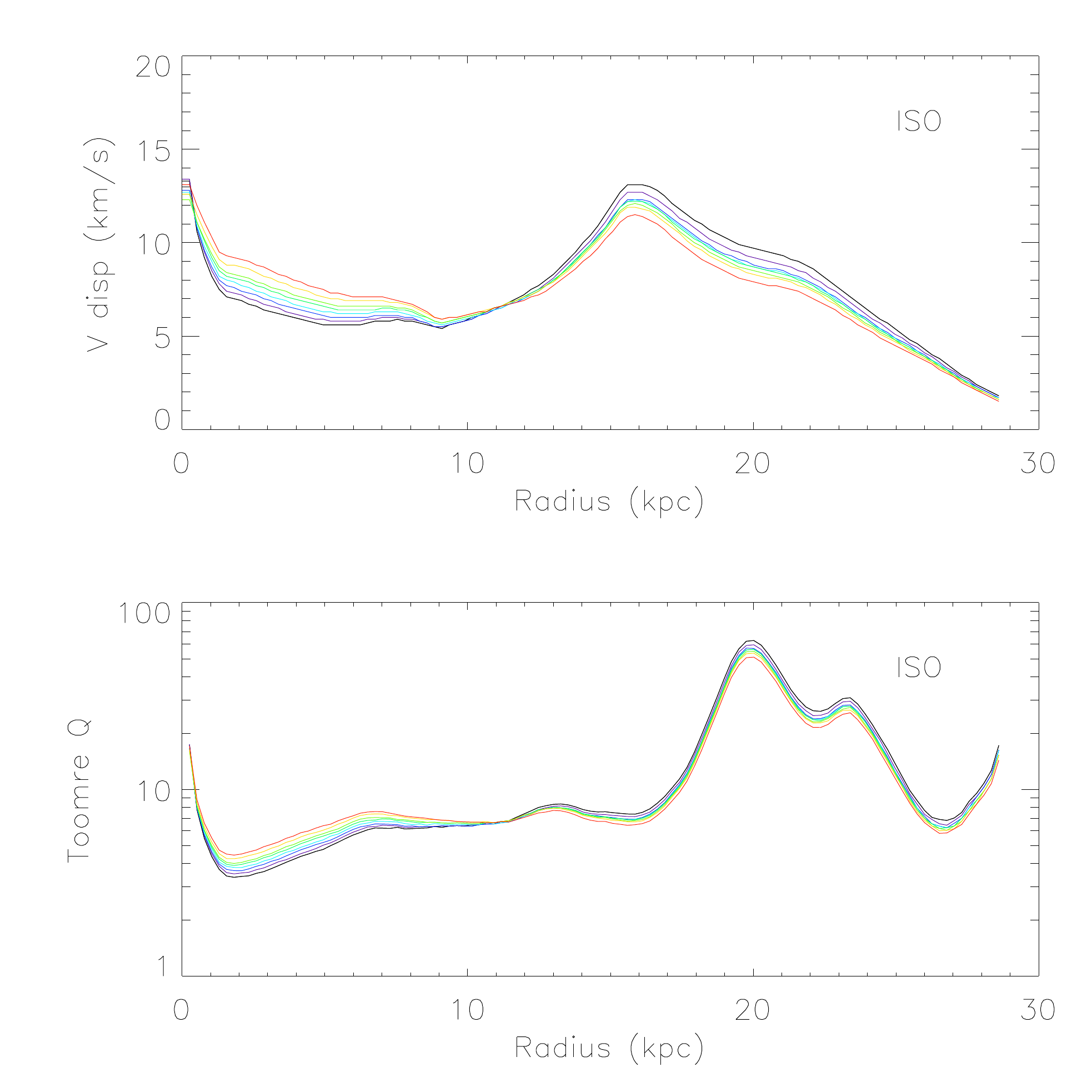}}
        \caption{Upper panel: vertical velocity dispersion of the gas halo of NGC~2683.
          Lower panel: derived Toomre $Q$ parameter.
          NFW halo and isothermal halo (ISO).
          The colors indicate different K' band mass-to-light ratios from blue (0.5) to red (1.3).
        } \label{fig:Vturb2683}
\end{figure}
We can go another step forward and calculate the velocity dispersion of the gas within the disk assuming that it is 
vertically sustained by turbulent motions. 
If vertical pressure equilibrium is assumed (see, e.g., Ostriker et al. 2010):
\begin{equation}
\rho v_{\rm turb}^{2}=\frac{1}{2} \pi G \Sigma \big(\Sigma (1+\frac{\rho_{\rm DM}}{\rho})+\Sigma_{*} \frac{v_{\rm turb}}
{v_{\rm disp}^{*}}\big)\ ,
\label{eq:velo}
\end{equation}
where $v_{\rm turb}$ is the vertical turbulent velocity of the gas, 
$G$ the gravitation constant, $\Sigma=\rho\,H$ the
gas surface density, $\rho$ the gas density, $\rho_{\rm DM}$ the dark matter density, $\Sigma_{*}$ the
stellar surface density, and $v_{\rm disp}^{*}$ the stellar vertical dispersion velocity.
The dark matter density is derived from the decomposition of the rotation curve (Fig.~\ref{fig:rotcurve}) and
$\rho_{\rm DM}=(v_{\rm DM}/R)^2/(4 \pi G)$. As before, we used K' band mass-to-light ratios
from $0.5$ to $1.3$. The stellar velocity dispersion is calculated assuming a constant vertical scale height of the stellar disk 
(Eq.~B3 of Leroy et al. 2008), which is $1/7.3$ times the radial scale length (Kregel et al. 2002).
Since the uncertainty of this fraction ($1/(7.3 \pm 2.2)$) resembles that of the
K' band mass-to-light ratio, we estimate the uncertainty of the velocity dispersion to be $\Delta v_{\rm turb} \sim 2$~km\,s$^{-1}$.

The derived vertical velocity dispersion of the gas decreases monotonically with increasing radius
from $10$~km\,s$^{-1}$ in the inner disk ($R=1$~kpc) to $7$~km\,s$^{-1}$ at $R=10$~kpc (Fig.~\ref{fig:Vturb2683}).
This is a typical behavior for nearby spiral galaxies (Tamburro et al. 2009, Fraternali et al. 2002).
Within the flare ($R=12$~kpc and $16$~kpc), it increases to $12$~km\,s$^{-1}$ and then decreases monotonically.
We believe that the increase is real, but its exact shape is mostly caused by our choice of the flare profile 
(Fig.~\ref{fig:flare}), which leads to a satisfactory reproduction of our observations.

\subsection{The Toomre $Q$ parameter \label{sec:toomre}}

The Toomre stability parameter is
\begin{equation}
Q=\frac{v_{\rm turb} \Omega}{\pi G \Sigma_{\rm tot}}\ ,
\end{equation}
where $\Sigma_{\rm tot}$ is the total gas surface density (see Sect.~\ref{sec:3D}), and $\Omega=v_{\rm rot}/R$.
The Toomre parameter is about $Q=4$-$7$ for radii smaller than $10$~kpc.
It then increases slightly to $Q=8$ at $13$~kpc, stays constant until $R=17$~kpc, and then rises steeply 
to a maximum of $Q \sim 60$ at $20$~kpc. The H{\sc i} ring at the extremity of the gas disk shows
a relatively low Toomre parameter of $Q \sim 6$.
With such high $Q$ values, NGC~2683 is similar to NGC~3351 and NGC~2841 (Leroy et al. 2008).

\subsection{The missing H{\sc i} halo}

NGC~2683 does not host an H{\sc i} halo around its optical disk and thin H{\sc i} disk.
This is consistent with the flocculent appearance of its disk  on public HST images.
Why does another equally massive spiral galaxy ($v_{\rm rot} \sim 200$~km\,s$^{-1}$), NGC~891, host an extended H{\sc i} halo, 
whereas we can exclude the existence of such a gas halo for NGC~2683?
The absence of a large-scale extended gaseous halo in the massive edge-on galaxy
NGC~4565 (Heald et al. 2011) can be explained by its much lower energy injection rate compared to NGC~891.
Is this also the case for NGC~2683?

Within the optical disk ($R < 9'$) of spiral galaxies, the thickness of gas disks can be explained well by energy injection
via supernovae explosions (galactic fountains; Shapiro \& Field 1976, Bregman 1980) and the vertical restoring force.
This force mainly depends on the disk mass surface density (gas and stars, see Eq.~\ref{eq:velo}).
Because NGC~2683 has a lower stellar mass than NGC~891, we do not expect its disk restoring force to be higher
than that of NGC~891. The energy injection rate is proportional to the local star formation rate  
$\Delta E/(\Delta t \Delta A) \propto \dot{\Sigma}_{*}$.

The star formation rate based on FIR emission of NGC~891 is $3.8$~M$_{\odot}$yr$^{-1}$ (Popescu et al. 2004).
NGC~2683's star formation rate based on radio continuum emission of $0.8$~M$_{\odot}$yr$^{-1}$ (Irwin et al. 1999)
is thus about five times lower than that of NGC~891. 
On the other hand, the flux ratios between NGC~891 and NGC~2683 in the IRAS $100$~$\mu$m band (Sanders et al. 2003) and 
Spitzer MIPS $70$~$\mu$m and $160$~$\mu$m bands are found to be around $5.5$. Assuming distances of $9.5$~Mpc and 
$7.7$~Mpc for NGC~891 and NGC~2683, respectively, we infer the star formation of NGC~2683 to be about eight times lower than that of NGC~891.

\begin{figure}
        \resizebox{\hsize}{!}{\includegraphics{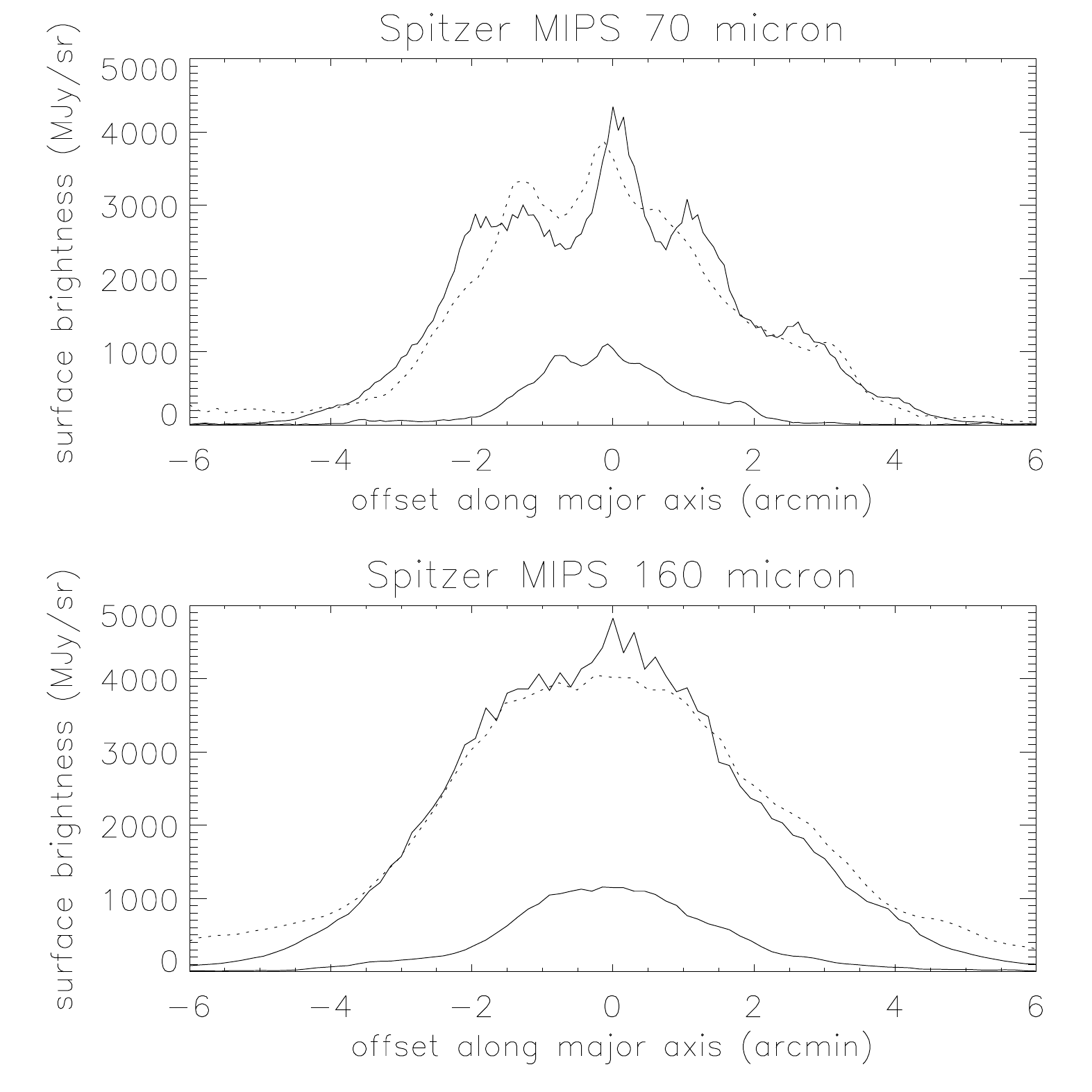}}
        \caption{Spitzer MIPS $70$~$\mu$m and $160$~$\mu$m surface brightness profiles of NGC~891 (upper solid lines)
          and NGC~2683 (lower solid lines) along the major axis summed over the minor axis.
          Dotted lines: profiles of NGC~2683 stretched by a factor $1.7$ and multiplied by a factor $3.5$. 
        } \label{fig:sfrspitzer}
\end{figure}

To compare the local energy injection of the two galaxies, we show the Spitzer MIPS $70$~$\mu$m and $160$~$\mu$m 
surface brightness profiles along the galaxies' major axis summed over the minor axis in Fig.~\ref{fig:sfrspitzer}.
NGC~2683 has a star forming disk that is about $1.7 \times \frac{9.5~{\rm Mpc}}{7.7~{\rm Mpc}}=2$
times less extended than NGC~891.
This is expected based on the smaller scale length of the stellar disk ($1.5$~kpc compared to $4$~kpc for NGC~891; 
Schechtman-Rook \& Bershady 2013). The nearly $3.5$ times lower surface brightness of the edge-on profile translates 
into a $3.5/2.0=1.8$ times lower local star formation rate or energy injection rate of NGC~2683
than for NGC~891. In the innermost disk where the local star formation rate is high, the deep galactic 
gravitational potential probably prevents the formation of a significant H{\sc i} halo.

We conclude that the small extent of the star forming disk
and the low local star formation rate of NGC~2683 explains the absence of an extended H{\sc i} halo around the optical disk.
Indeed, the H{\sc i} halo of NGC~891 is the most prominent at galactic radii between $2.5$ and $15$~kpc (Fig.~13 of
Oosterloo et al. 2007), whereas the star forming disk of NGC~2983 only extends to $\sim 3$~kpc.

\subsection{How to sustain the flaring H{\sc i} disk}

NGC~2683's gas disk is flaring so is very thick at radii well beyond the optical radius where the rate of star formation is 
very low or even nonexistent. A galactic-fountain origin 
for the flare gas beyond the optical radius is thus excluded.
For the majority of the flare region, the gas velocity dispersion is well above the speed of sound ($\sim 6$~km\,s$^{-1}$).
To maintain the gas velocity dispersion at this high level, four physical mechanisms can be suggested:
\begin{itemize}
\item
energy injection by supernovae,
\item
magneto-rotational instabilities (Balbus \& Hawley 1991),
\item
ISM stirring by dark matter substructure,
\item
external gas accretion.
\end{itemize}
In the following we compare the energy injection rates of these mechanisms with the energy
dissipation rate via turbulence.

Energy injection by supernovae:
Following Vollmer \& Beckert (2003) energy flux conservation between the energy injection via
supernovae and the energy dissipation rate via turbulence is
\begin{equation}
\Sigma \nu \frac{v_{\rm turb}^2}{l_{\rm driv}^{2}}=\xi \dot{\Sigma}_{*}\ ,
\end{equation}
where $\nu=v_{\rm turb} l_{\rm driv}$ is the gas viscosity, $l_{\rm driv}$ the turbulent
driving length scale, and $\xi=3.6 \times 10^{-8}$~pc$^{2}$yr$^{-2}$ is normalized by
the Galactic supernova rate.
For a driving length scale of $l_{\rm driv}=1$~kpc, a turbulent velocity of $v_{\rm turb}=8$~km\,s$^{-1}$,
and a gas surface density of $\Sigma=0.5$~M$_{\odot}$pc$^{-2}$, we obtain
$\dot{\Sigma}_{*}=8 \times 10^{-12}$~M$_{\odot}$yr$^{-1}$pc$^{-2}$. Such levels of star formation 
are observed in the outer disks of nearby spiral galaxies (Bigiel et al. 2010).
The energy injection rate is $\Delta E_{*}/(\Delta A\ \Delta t) \sim 2 \times 10^{-8}$~erg\,cm$^{-2}$s$^{-1}$.

Magneto-rotational instabilities:
Following Tamburro et al. (2009), the energy injection rate by magneto-rotational instabilities 
for a disk height of 1~kpc, a magnetic field strength of 1~$\mu$G, and an angular velocity
of $\Omega=10^{-8}$~yr$^{-1}$ are 
$\Delta E_{\rm MRI}/(\Delta A\ \Delta t) \sim 2 \times 10^{-8}$~erg\,cm$^{-2}$s$^{-1}$.
The energy injection due to magneto-rotational instabilities is thus comparable to the one caused by supernovae.

ISM stirring by dark matter substructure:
For the stirring of the gas disk by the dark matter substructure, we can estimate the
energy deposited by a dark matter subhalo crossing the disk is $\Delta E \sim \epsilon M_{\rm halo} v^{2}$.
The timescale of interactions between a halo of mean mass $M_{\rm halo}$ is $t_{\rm coll}$.
The energy injection rate is thus
\begin{equation}
\frac{\Delta E}{\Delta A \Delta t}=\frac{N \epsilon M_{\rm halo} v^{2}}{\pi R^{2} t_{\rm coll}}\ . 
\end{equation}
We assume $v=150$~km\,s$^{-1}$, $M_{\rm halo}=10^{7}$~M$_{\odot}$, $R=20$~kpc, and a 
collision timescale $t_{\rm coll} \sim \pi R / v \sim 4 \times 10^{8}$~yr. 
With $N \sim 20,$ a fraction of $\epsilon \sim 0.03$ of the halo's kinetic energy has to be injected into the ISM 
to balance the energy dissipation rate via turbulence.

External gas accretion:
In the case of external accretion, one can estimate the necessary infall velocity $v_{\rm infall}$
to maintain local gas turbulence via the following energy flux conservation equation:
\begin{equation}
\Sigma \frac{v_{\rm turb}^3}{H}=\frac{\dot{M}}{\Delta A} v_{\rm infall}^2 ,
\end{equation}
where $\dot{M} \sim 0.2$~M$_{\odot}$yr$^{-1}$ (Sancisi et al. 2008) is the external mass accretion rate, $\Sigma=0.5$~M$_{\odot}$pc$^{-2}$,
$H=3$~kpc, and $\Delta A \sim \pi (20~{\rm kpc})^{2}$.
With these numbers we obtain $v_{\rm infall} \sim 24$~km\,s$^{-1}$.
The local dissipation timescale of turbulence is close to the crossing time
$t_{\rm cross} \sim H/v_{\rm turb} \sim 3 \times 10^{8}$~yr (Mac Low et al. 1999).
The timescale of external accretion is $t_{\rm acc} \sim M_{\rm outer HI}/\dot{M} \sim 3$~Gyr, 
where $M_{\rm outer HI} \sim 6 \times 10^{8}$~M$_{\odot}$ is the mass of the flaring part of the gas disk.
Thus, it is only possible to maintain ISM turbulence by external gas accretion if the local dissipation timescale of turbulence
is decreased by a factor of about ten with respect to its value for fully developed turbulence in a star forming disk. 

For accreting thick gas tori around active
galactic nuclei,
Vollmer \& Davies (2013) suggest that
massive and rapid gas accretion can lead to adiabatic compression of the ISM. This leads to (i) an increase in the
gas velocity dispersion, (ii) intermittent turbulence, and (iii) quenching of star formation.
Vollmer \& Davies (2013) propose that turbulent adiabatic compression can lead to intermittent turbulence and a
subsequent decrease in the local dissipation timescale of turbulence (by the area filling factor of dense clumps or clouds). 
This mechanism might also act in the flaring gas disk of NGC~2683.

From these estimates none of the suggested mechanisms, which might be responsible for the
driving of ISM turbulence, can be discarded. All effects might contribute to setting the overall
velocity dispersion, but locally one of the four mechanisms might dominate according to the local conditions.
The existence of the complex large-scale warping (Sect.~\ref{sec:3D}) with asymmetries (Sect.~\ref{sec:atomic}) 
and small-scale structure (Sect.~\ref{sec:hidistribution}) might indicate that 
large scale external gas accretion has occurred and/or is occurring in NGC~2683.
Alternatively, a triaxial dark matter halo can also induce a large-scale warp (e.g., Binney 1992).
Since H{\sc i} distributions of most spiral galaxies, which extend beyond the optical disk, show warps (Garcia-Ruiz et al. 2002) and
are flared (O'Brien et al. 2010), we propose the following scenario for NGC~2683:
\begin{itemize}
\item
Recent external gas accretion (within the last few Gyr) has added the atomic gas beyond the optical diameter (see van der Kruit 2007).
\item
Since the angular momentum of the infalling gas is not expected to be the same as that of the disk, a kinetic and
spatial warp forms.
\item
During circularization, gas is compressed in colliding rotating gas streams/arms.
\item
If the compression is adiabatic, i.e. it increases the gas velocity dispersion, ISM turbulence becomes intermittent
and star formation is quenched. In this case gas compression feeds and maintains the ISM turbulence.
\item
If the compressed turbulent gas can dissipate the injected mechanical energy, star formation proceeds in the compression 
region and the energy injection by subsequent SN explosions feeds and maintains ISM turbulence.
\item
Since we expect equipartition between the turbulent kinetic energy and the magnetic field energy, 
magneto-rotational instabilities might become important in regions where turbulence has enhanced the magnetic field.
\end{itemize}
This rather speculative scenario might be valid for other spiral galaxies with extended low surface density atomic gas.
In addition, a triaxial halo might induce and/or modify a warp of the outer H{\sc i} disk.

\section{Conclusions \label{sec:conclusions}}

New deep VLA D array H{\sc i} observations of the nearby highly inclined spiral galaxy NGC~2683 have been presented.
Archival C array data were processed and added to the new observations.
In the D array data an rms noise level of 1.0~mJy/beam is reached in a 5.16~km\,s$^{-1}$ channel.
The total H{\sc i} mass of NGC~2683 is $M_{\rm HI}=1.42 \times 10^{9}$~M$_{\odot}$.
As already shown by Casertano \& van Gorkom (1991), the gas disk extends up to about three times the optical radius. At the extremity of the H{\sc i} disk ($\sim 27$~kpc), we observe a ring-like structure whose projected 
surface density distribution has a blob-like structure.
The position velocity diagram along the major axis does not show any counter-rotating gas.

To investigate the 3D structure of the atomic gas disk, we made different 3D models for which 
we produced model H{\sc i} data cubes. These models have the following main components: 
(i) a thin gas disk with a thickness of $500$~pc, 
(ii) different gas flares at galactic radii larger than $9$~kpc (Fig.~\ref{fig:flare}), (iii) a possible warp of the disk, and
(iii) an outer gas ring ($R > 25$~kpc).

The main ingredients of our best-fit model are: 
(i) a thin disk inclined by $80^{\circ}$, (ii) a crude approximation of spiral and/or bar structure by
an elliptical surface density distribution of the gas disk,
(iii) a slight warp in inclination between $10$~kpc$ \leq R \leq 20$~kpc,
(iv) an exponential flare that rises from $0.5$~kpc at $R=9$~kpc 
to $4$~kpc at $R=15$~kpc, stays constant until $R=22$~kpc, and decreases its height for $R > 22$~kpc
(flare F3 in Fig.~\ref{fig:flare}), and (v) 
a low surface density gas ring with a vertical offset of $1.3$~kpc.

The slope of NGC~2683's flare is comparable, but somewhat steeper than those of other spiral galaxies. 
NGC~2683's maximum height of the flare is also comparable with those of other galaxies. On the other hand,
a saturation of the flare is only observed in NGC~2683.

Based on the comparison between the high resolution model and observations, we excluded that there is 
an extended atomic gas halo around the optical and thin gas disk (Figs.~\ref{fig:PV2HR} and \ref{fig:canauxHR1}). 
By comparing the disk properties (local star formation rate and mass surface density) of NGC~2683 with that of NGC~891, 
we concluded that the small extent of the star forming disk
and the low local star formation rate of NGC~2683 might explain the absence of an extended H{\sc i} halo around the optical disk.

Under the assumption of vertical hydrostatic equilibrium, we derived
the vertical velocity dispersion of the gas. As in other nearby galaxies
(Tamburro et al. 2009, Fraternali et al. 2002), the velocity dispersion decreases monotonically with increasing radius within 
the optical disk. The velocity dispersion in the inner disk is about $8$~km\,s$^{-1}$.

Since the gas disk flares, i.e. is considerably thick, beyond the optical radius where star formation is low
or even absent, it is improbable that galactic fountains are responsible for the
high velocity dispersion (exceeding the thermal velocity dispersion). We estimated the energy dissipation rate due to turbulence and
compared it to the energy injection rates due to (i) supernova explosions, (ii) magneto-rotational
instabilities, (iii) ISM stirring by dark matter substructure, and (iv) external gas accretion. 
It is found that none of these mechanisms can be definitely discarded.

The existence of the complex large-scale warping (Sect.~\ref{sec:3D}) and asymmetries (Sect.~\ref{sec:atomic}) indicates that 
large scale external gas accretion has occurred and/or is occurring in NGC~2683.
Since most galaxies with H{\sc i} distributions extending beyond the optical disk show warps (Garcia-Ruiz et al. 2002) and
are flared (O'Brien et al. 2010), we propose a scenario where recent external gas accretion (within the last view Gyr) 
has added the atomic gas beyond the optical radius, in which a kinetic and spatial warp forms.
During circularization, gas is compressed in colliding rotating gas streams/arms.
If the compression is adiabatic, i.e. it increases the gas velocity dispersion, ISM turbulence becomes intermittent
and star formation is expected to be quenched. In this case gas compression feeds and maintains the ISM turbulence.
If the compressed turbulent gas can dissipate the injected mechanical energy, star formation proceeds in the compression 
region, and the energy injection by subsequent SN explosions feeds and maintains ISM turbulence.

\begin{acknowledgements}
We would like to thank the anonymous referee for constructive comments that helped to improve this article significantly. 
\end{acknowledgements}

\begin{appendix}

\section{H{\sc i} position velocity diagrams \label{app1}}

\begin{figure*}
        \resizebox{15cm}{!}{\includegraphics{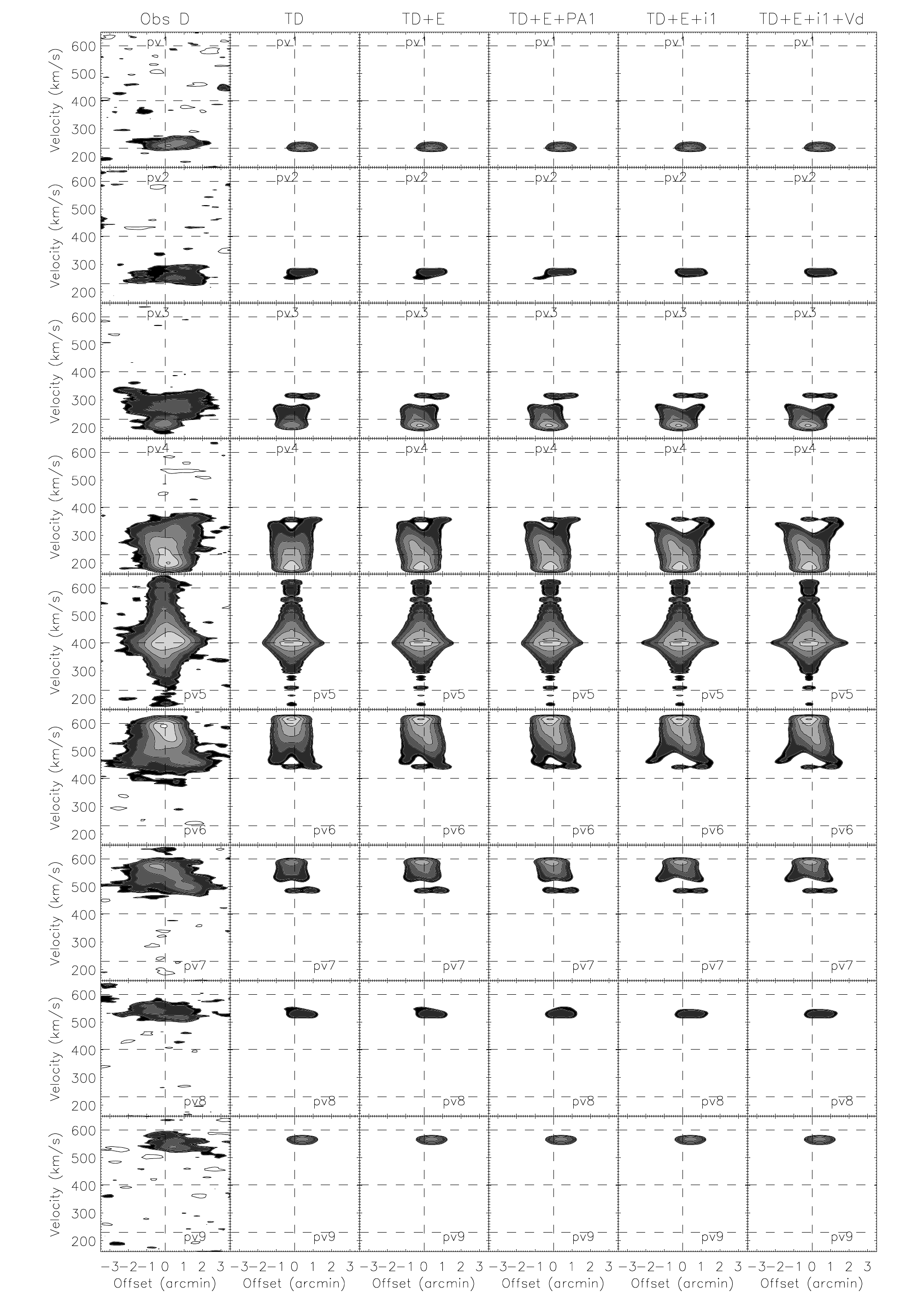}}
        \caption{NGC~2683 H{\sc i} D array and model position velocity diagrams.
          The model ingredients are a thin disk (TD), an elliptical component (E), a warp in position angle (PA),
          a warp in inclination (i), and a radially decreasing velocity dispersion (Vd).
          The contour levels are $(-2,2,3,6,12,24,48,96) \times 0.7$~mJy/beam.
          The resolution is $61'' \times 51''$.
        } \label{fig:PV1BR}
\end{figure*}

\begin{figure*}
        \resizebox{15cm}{!}{\includegraphics{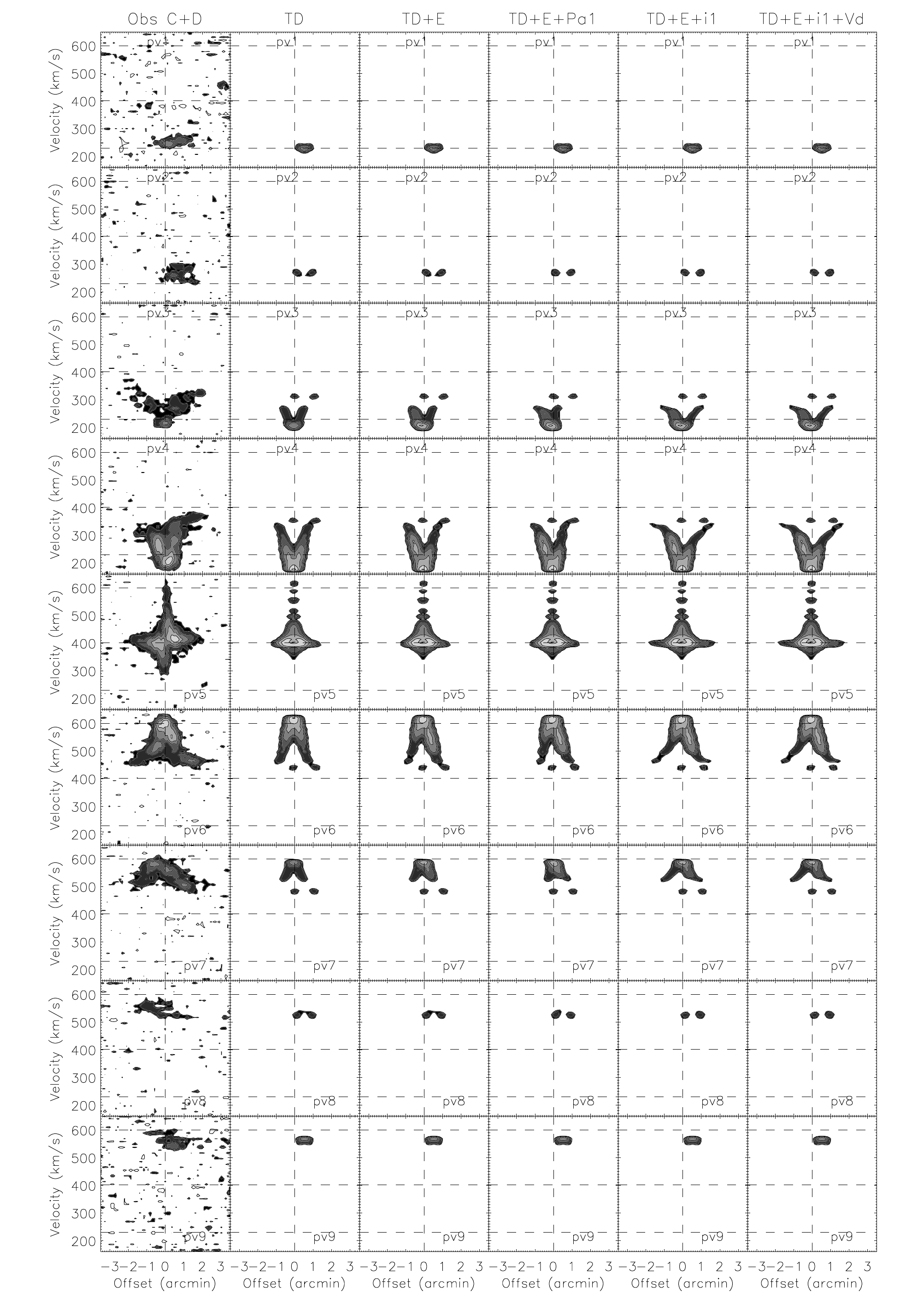}}
        \caption{NGC~2683 H{\sc i} C+D array and model position velocity diagrams.
          The model ingredients are a thin disk (TD), an elliptical component (E), a warp in position angle (PA),
          a warp in inclination (i), and a radially decreasing velocity dispersion (Vd).
          The contour levels are $(-2,2,3,6,12,24,48,96) \times 0.3$~mJy/beam.
          The resolution is $19'' \times 18''$.
        } \label{fig:PV1HR}
\end{figure*}

\begin{figure*}
        \resizebox{15cm}{!}{\includegraphics{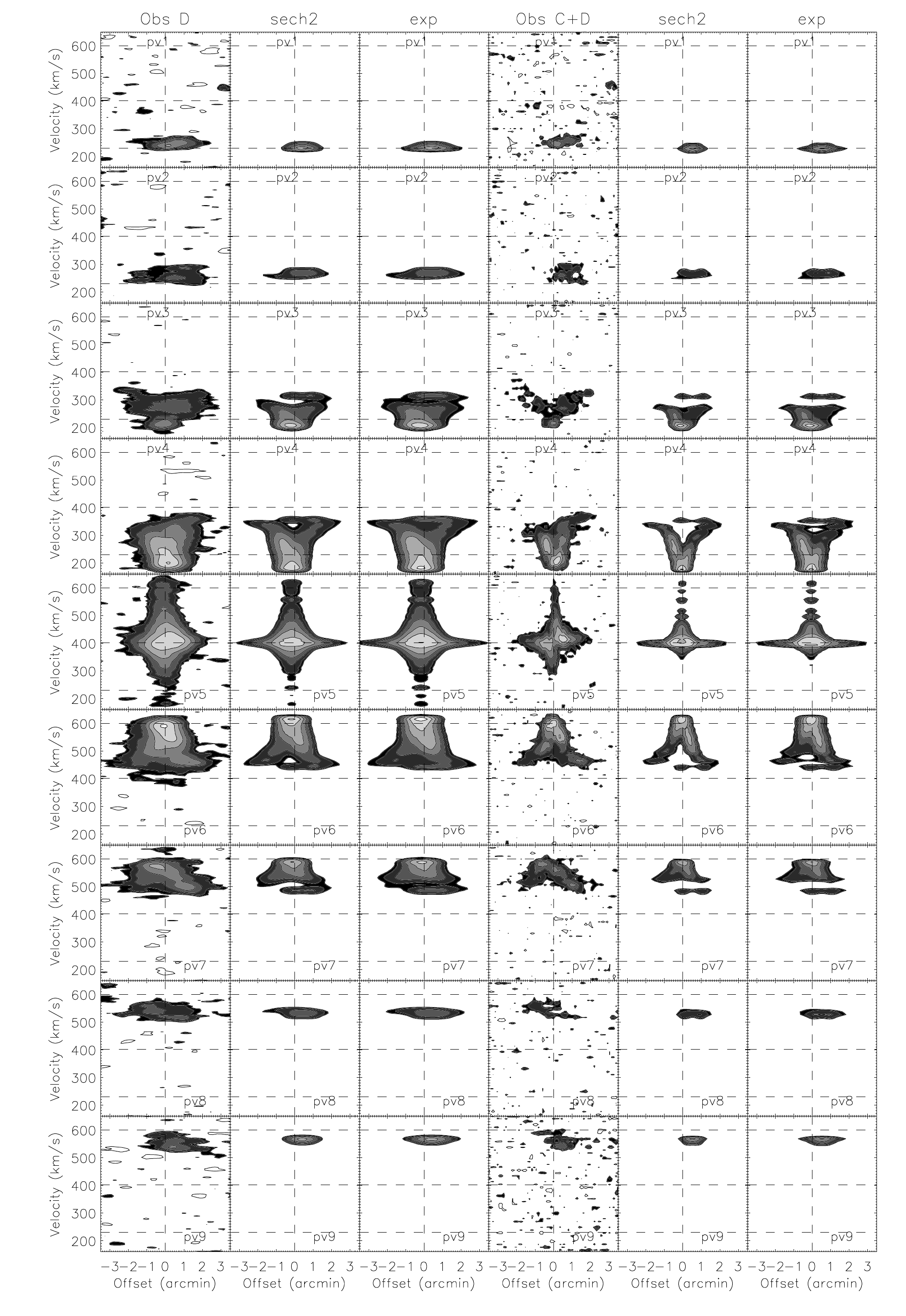}}
        \caption{Models with different vertical structures of the gas disk (${\rm sech}^2$ and exponential).
          Left panels:  NGC~2683 H{\sc i} D array and model position velocity diagrams.
          The contour levels are $(-2,2,3,6,12,24,48,96) \times 0.7$~mJy/beam.
          The resolution is $61'' \times 51''$. Models include a flare with ${\rm sech}^{2}$ and exponential vertical profiles.
          Right panels: NGC~2683 H{\sc i} C+D array and model position velocity diagrams.
          The contour levels are $(-2,2,3,6,12,24,48,96) \times 0.3$~mJy/beam.
          The resolution is $19'' \times 18''$. Models include a flare with ${\rm sech}^{2}$ and exponential vertical profiles.
        }\label{fig:PV4BR}
\end{figure*} 

\section{H{\sc i} channel maps \label{app}}

\begin{figure*}
        \resizebox{\hsize}{!}{\includegraphics{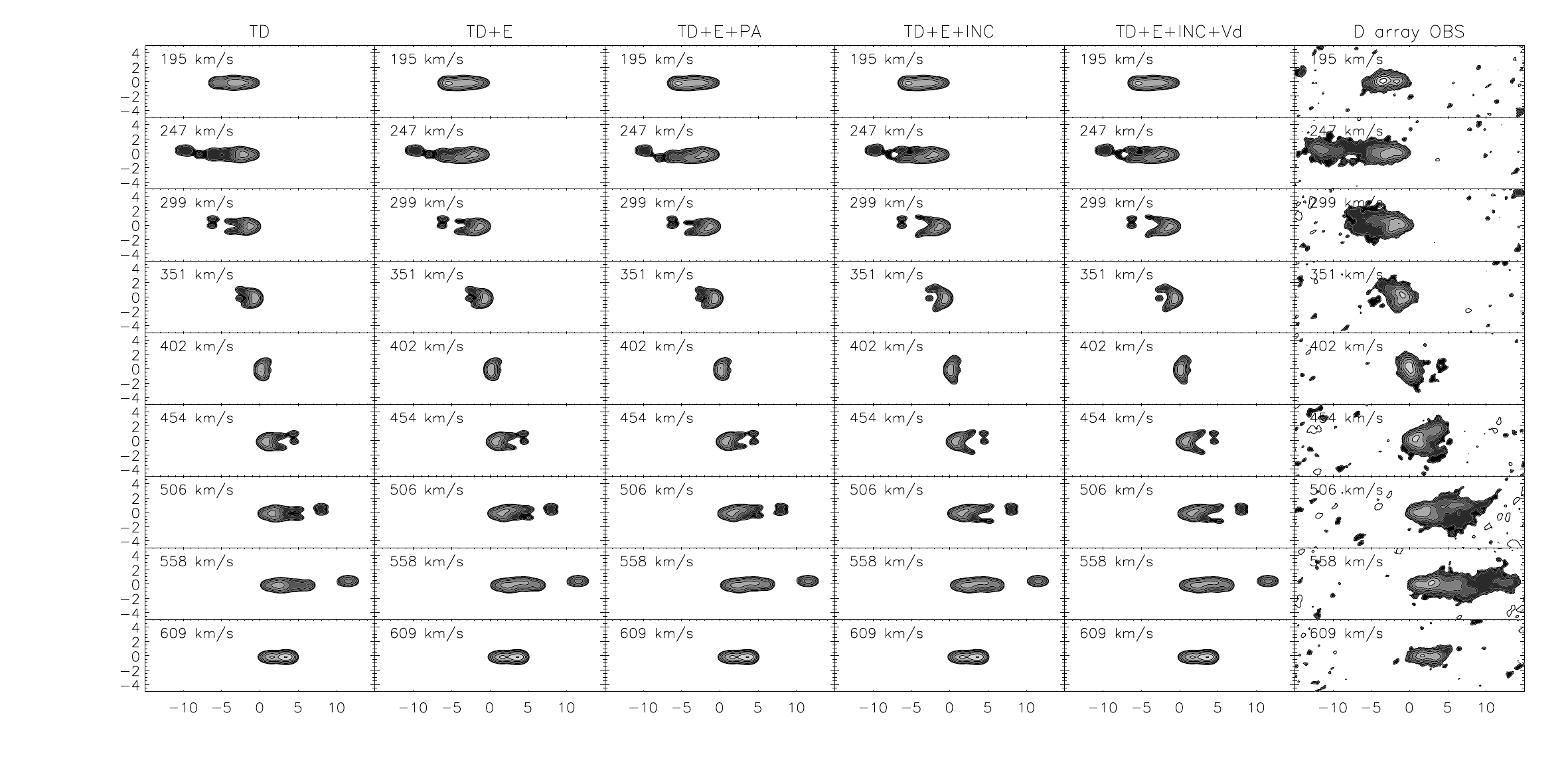}}
        \caption{Selected H{\sc i} D array and model channel maps.
          The model ingredients are a thin disk (TD), an elliptical component (E), a warp in position angle (PA),
          a warp in inclination (INC), and a radially decreasing velocity dispersion (Vd).
          The offsets along the $x$- and $y$-axis are in arcmin.
          The contour levels are $(-2,2,3,6,12,24,48,96) \times 0.8$~mJy/beam.
          The resolution is $61'' \times 51''$. The galaxy was rotated by $45^{\circ}$.
        } \label{fig:canauxBR2}
\end{figure*} 
\begin{figure*}
        \resizebox{\hsize}{!}{\includegraphics{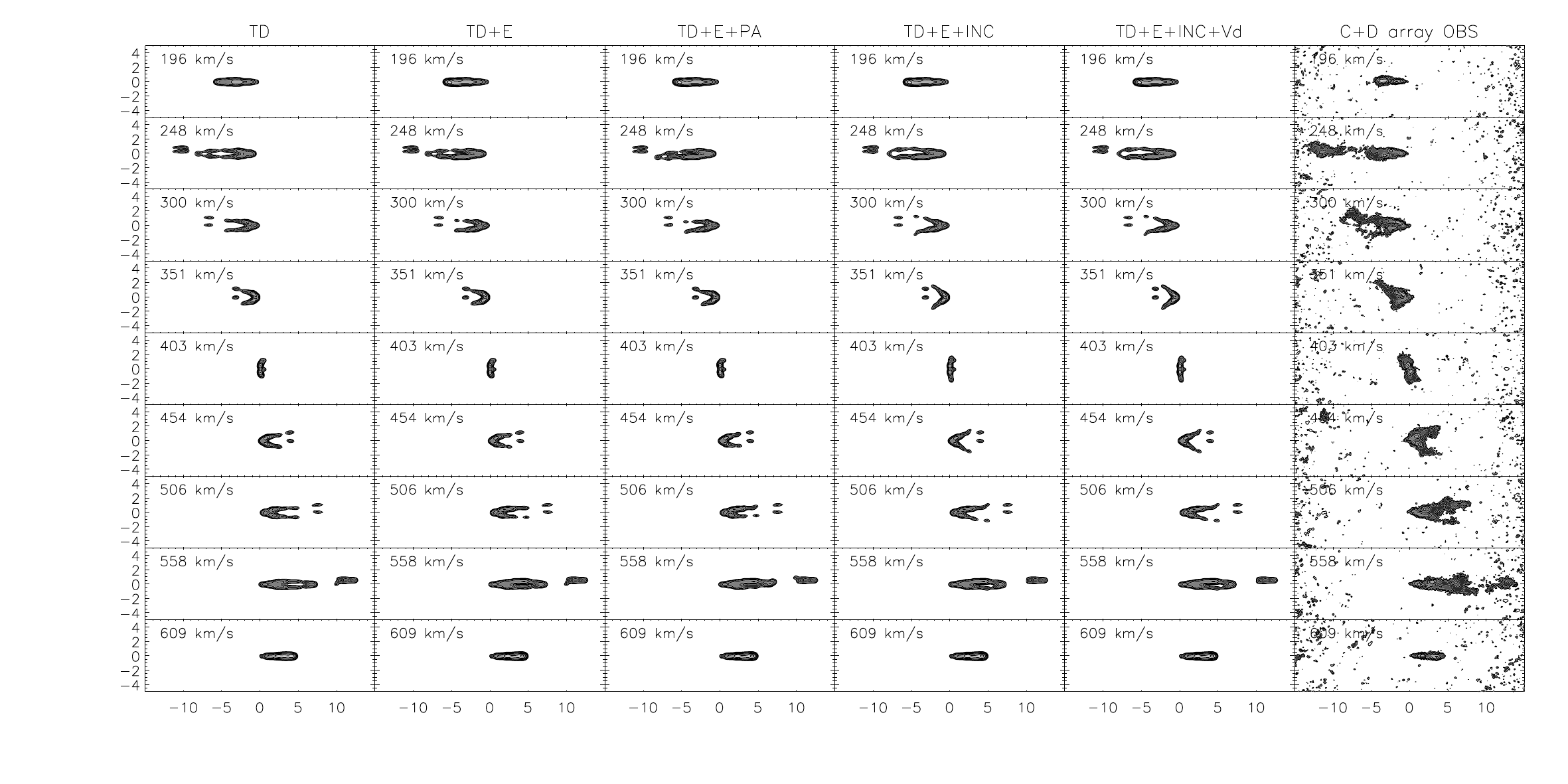}}
        \caption{Selected H{\sc i} C+D array and model channel maps.
          The model ingredients are a thin disk (TD), an elliptical component (E), a warp in position angle (PA),
          a warp in inclination (INC), and a radially decreasing velocity dispersion (Vd).
          The offsets along the $x$- and $y$-axis are in arcmin.
          The contour levels are $(-2,2,3,6,12,24,48,96) \times 0.4$~mJy/beam.
          The resolution is $19'' \times 18''$. The galaxy was rotated by $45^{\circ}$.
        } \label{fig:canauxHR2}
\end{figure*}

\begin{figure*}
        \resizebox{\hsize}{!}{\includegraphics{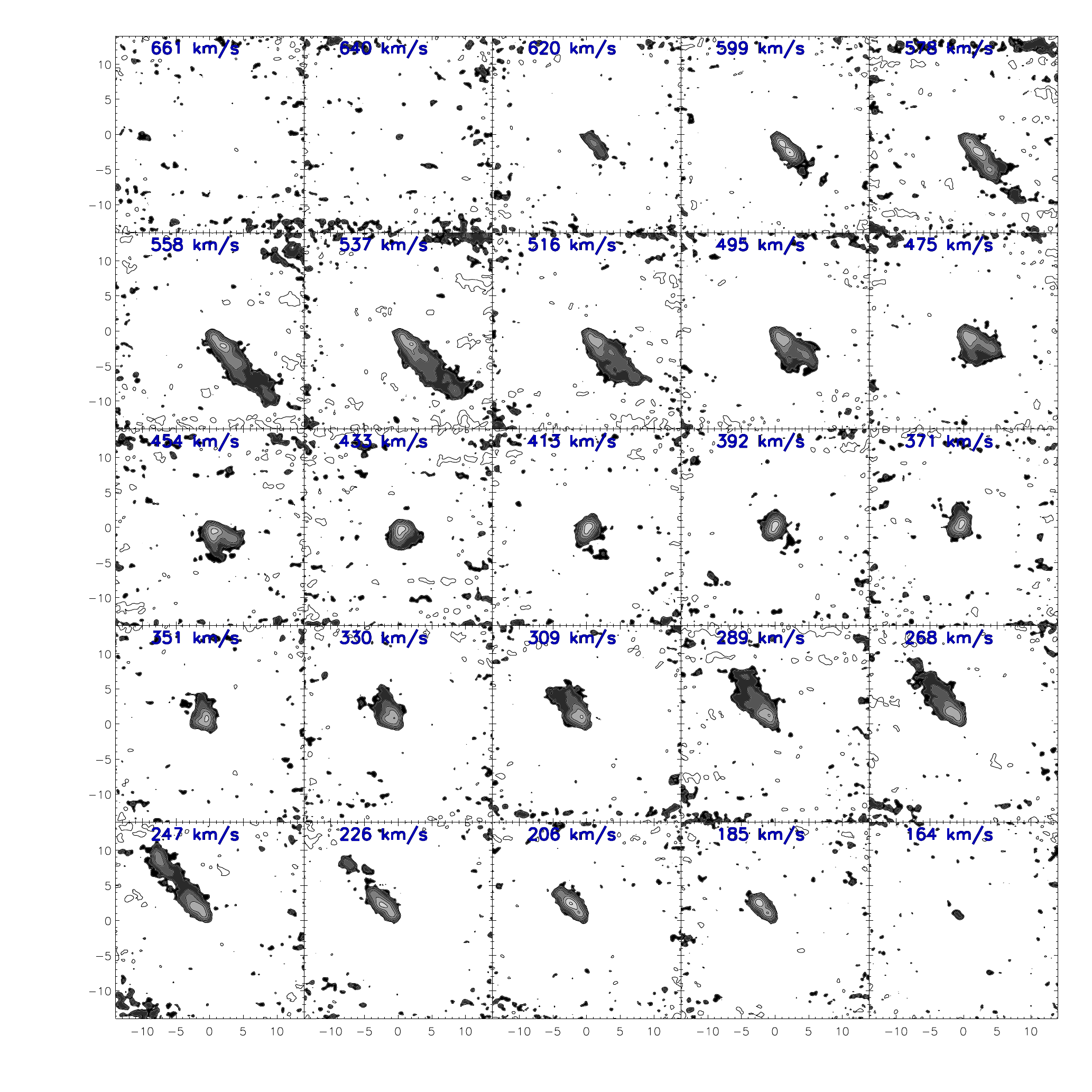}}
        \caption{NGC~2683 H{\sc i} D array channels maps.
          The contour levels are $(-2,2,3,6,12,24,48,96) \times 0.8$~mJy/beam.
          The resolution is $61'' \times 51''$.
          The offsets along the $x$- and $y$-axis are in arcmin.
        } \label{fig:channels_D}
\end{figure*}

\begin{figure*}
        \resizebox{\hsize}{!}{\includegraphics{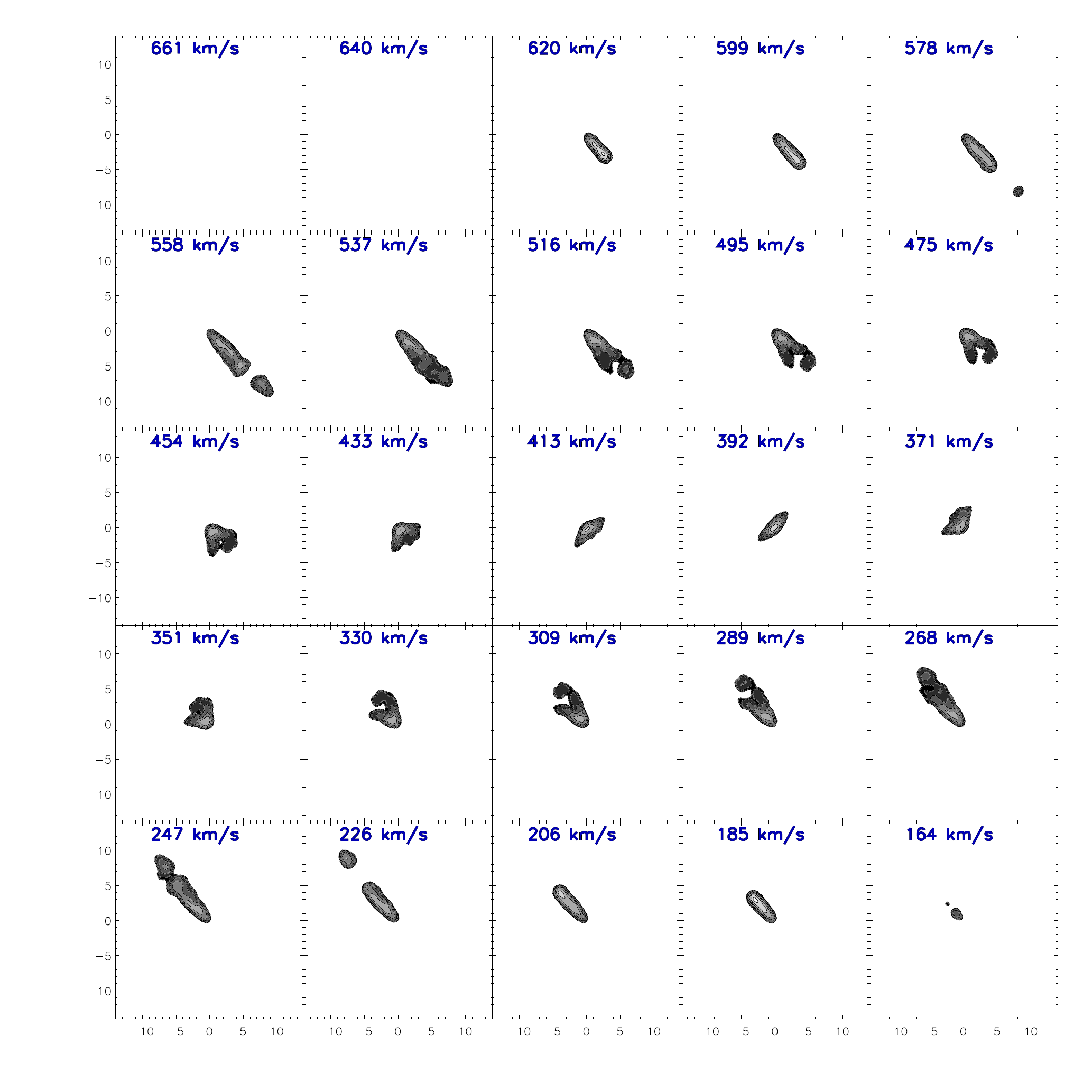}}
        \caption{Best-fit model channel maps.  
          The model ingredients are a thin disk (TD), an elliptical component (E), a warp in inclination (INC), 
          a radially decreasing velocity dispersion (Vd), and a flare (F3).
          The resolution is $61'' \times 51''$. The contour levels are $(-2,2,3,6,12,24,48,96) \times 0.8$~mJy/beam.
          The channels of the original data cube were averaged to yield a channel separation of $20$~km\,s$^{-1}$.
          The offsets along the $x$- and $y$-axes are in arcmin.
        } \label{fig:canauxbestBR}
\end{figure*}

\begin{figure*}
        \resizebox{\hsize}{!}{\includegraphics{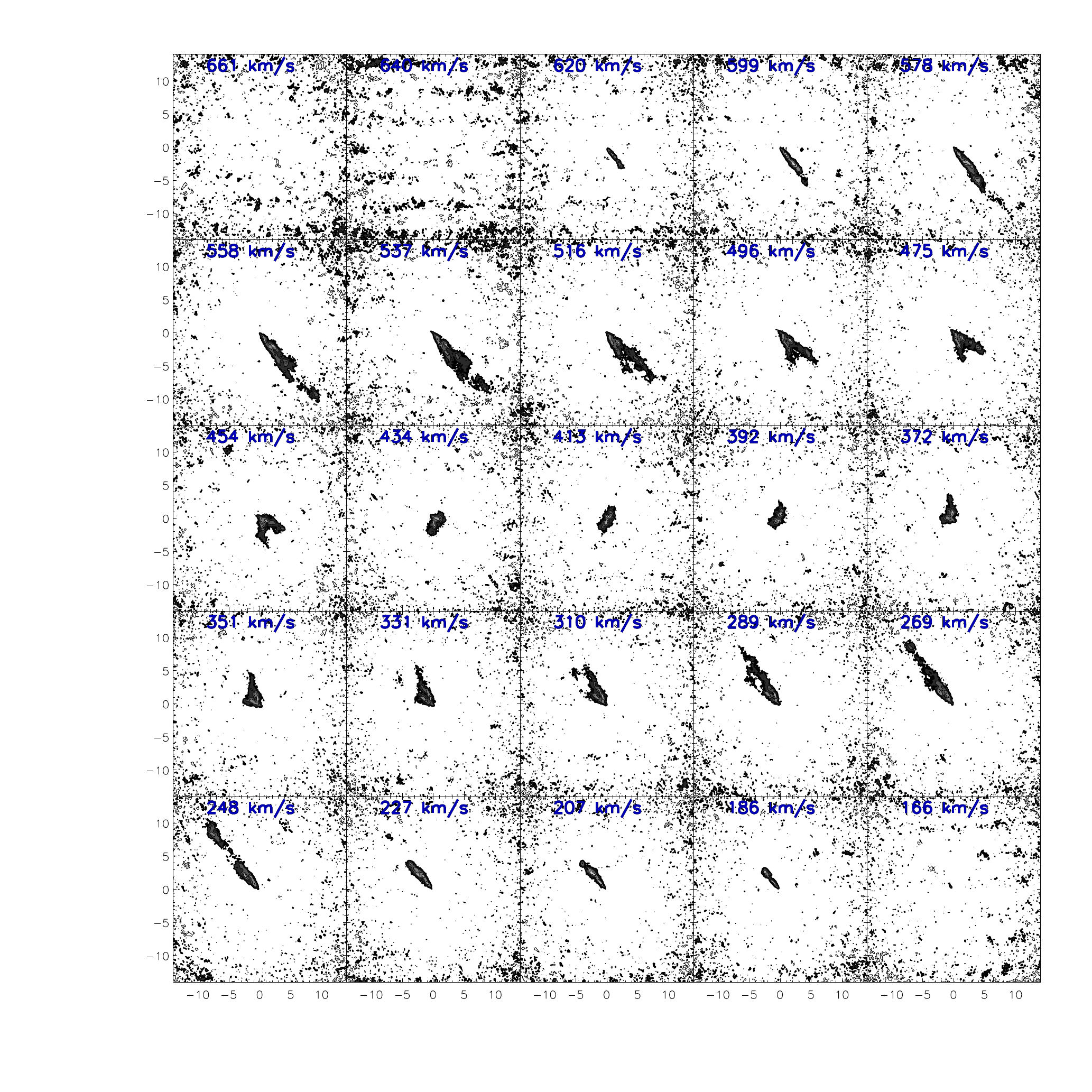}}
        \caption{NGC~2683 H{\sc i} C+D array channels maps.
          The contour levels are $(-2,2,3,6,12,24,48,96) \times 0.4$~mJy/beam.
          The resolution is $19'' \times 18''$.
          The offsets along the $x$- and $y$-axes are in arcmin.
        } \label{fig:channels_CD}
\end{figure*}

\begin{figure*}
        \resizebox{\hsize}{!}{\includegraphics{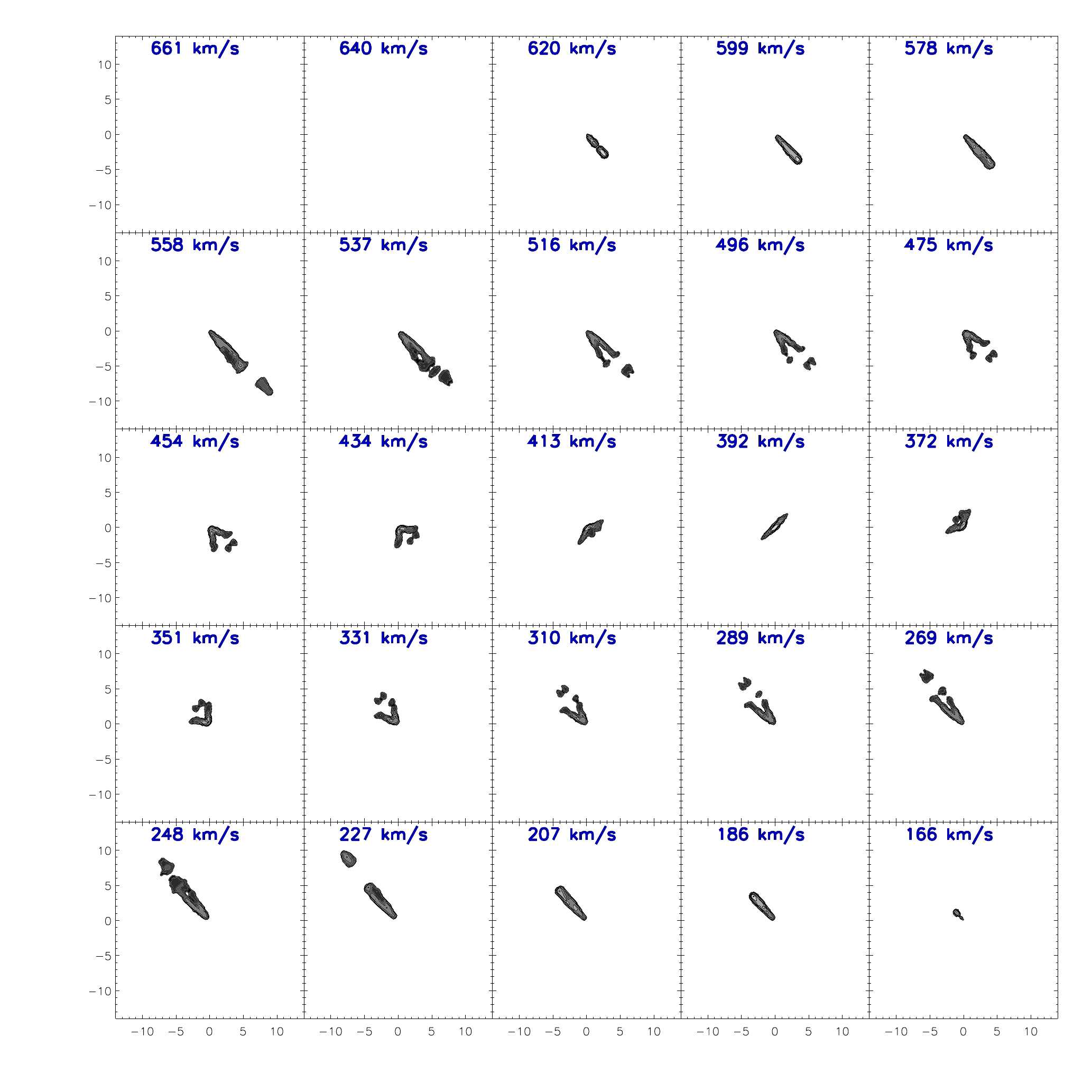}}
        \caption{Best-fit model channel maps.
          The model ingredients are a thin disk (TD), an elliptical component (E), a warp in inclination (INC), 
          a radially decreasing velocity dispersion (Vd), and a flare (F3).
          The contour levels are $(-2,2,3,6,12,24,48,96) \times 0.4$~mJy/beam.
          The resolution is $19'' \times 18''$.
          The offsets along the $x$- and $y$-axes are in arcmin.
        } \label{fig:canauxbestHR}
\end{figure*}

\end{appendix}

\end{document}